 \documentclass[final,3p,times]{elsarticle}

\usepackage{amsmath,amssymb,amsfonts}
\usepackage{amsthm}
\usepackage[dvipsnames,table]{xcolor}
\usepackage{colortbl}
\usepackage{dsfont}
\usepackage{bbm}

\usepackage[labelfont=bf]{caption}
\usepackage{placeins}

\usepackage{tikz}
\definecolor{orange}{HTML}{d9944f}
\usetikzlibrary{patterns,arrows,positioning,fit}
\usetikzlibrary{decorations.pathreplacing}
\usetikzlibrary{decorations.pathmorphing}

\usepackage{graphicx,subcaption}
\usepackage{url}
\usepackage{doi}
\usepackage{hyperref}
\usepackage[capitalize]{cleveref}
\usepackage{float}
\usepackage{orcidlink}
\usepackage{booktabs}
\usepackage{tabularx}

\usepackage{multirow}
\usepackage{multicol}

\newtheorem{theorem}{Theorem}[section]

\newtheorem{remark}[theorem]{Remark}

\usetikzlibrary{arrows.meta, positioning, calc}
\usepackage{nicematrix}

\usepackage{array}

\newcolumntype{L}[1]{>{\raggedright\arraybackslash}p{#1}}
\newcolumntype{D}{>{\hsize=2.0\hsize\raggedright\arraybackslash}X}
\newcolumntype{V}{>{\hsize=0.70\hsize\centering\arraybackslash}X}

\usepackage[figuresright]{rotating}
\usepackage{orcidlink}

\usepackage[ruled,vlined]{algorithm2e}
 \SetKwComment{Comment}{}{}
 \SetKwFor{OMPFOR}{omp for}{do}{end}
\SetKwFor{OMPSIMD}{omp simd}{do}{end}
 \let\oldnl\nl
\newcommand{\nonl}{\renewcommand{\nl}{\let\nl\oldnl}}
\makeatletter 
\renewcommand{\algocf@typo}{}
\makeatother

\usepackage{todonotes}
\newboolean{SHOWCOMMENTS}
\setboolean{SHOWCOMMENTS}{true}
\ifthenelse{\boolean{SHOWCOMMENTS}}{
\newcommand{\tdHZ}[1]
{\todo[color=green!25,inline]{\footnotesize{\bf Henrik:} #1}}
}{
\newcommand{\tdHZ}[1]{}
}

\ifthenelse{\boolean{SHOWCOMMENTS}}{
\newcommand{\tdMK}[1]
{\todo[color=blue!25,inline]{\footnotesize{\bf Martin:} #1}}
}{
\newcommand{\tdMK}[1]{}
}

\renewcommand{\d}[1]{\mathrm{d}#1}

\newcommand{\vismode}{showpapertwo}
\newcommand{\papertwo}[1]{%
  \ifx\vismode\showpapertwo
  \else
    {\color{blue} #1}%
  \fi
}

\begin{document}

\begin{frontmatter}




\title{High-performance metapopulation models for epidemic emergencies:\\ Scalability and performance of implicit and explicit mobility representations on~CPUs~and~GPUs}


\author[a]{Henrik Zunker \orcidlink{0000-0002-9825-365X}} 
\ead{Henrik.Zunker@DLR.de}
\author[a]{Carlotta Gerstein \orcidlink{0009-0004-4410-0502}}
\author[a,b]{Martin J. Kühn\orcidlink{0000-0002-0906-6984}\corref{cor1}}
\cortext[cor1]{Corresponding author}
\ead{Martin.Kuehn@DLR.de}

\address[a]{Institute of Software Technology, Department of High-Performance Computing,
German Aerospace Center, Cologne, Germany}
\address[b]{Bonn Center for Mathematical Life Sciences and Life and Medical Sciences Institute, University of Bonn, Germany}

\begin{abstract}
Epidemics form a particular subclass of natural disasters impacting countries or entire continents, where the most suitable reaction might differ substantially across the spatio-temporal dimensions. Spatially resolved models based on ordinary differential equations (ODEs) are an efficient modeling approach for large and distributed populations. They are commonly referred to as metapopulation models and allow a realistic representation of spatial heterogeneities and mobility. Yet, their application in epidemic settings, where high time pressure prevails, requires highly efficient and parallelized implementations leveraging the potential of modern supercomputers.

The different ways to represent mobility in metapopulation models affect not only epidemic trajectories but also the computational cost of the simulations. In contrast to implicit mobility modeling, where populations are coupled through transition and transmission intensities, explicit Lagrangian mobility tracks traveling subpopulations according to their home and current location. 

In this paper, we compare metapopulation models with implicit and explicit mobility. A four-patch experiment further shows that the epidemic trajectories of the two formulations remain close, but the differences increase with the mobility cycle length. We present CPU- and GPU-based parallelizations and analyze fully connected networks with up to 16,384 patches on 128 CPU cores as well as NVIDIA A100 and H100 GPUs. Roofline analyses show close to maximum performance for the most time-consuming GPU subroutines and identify data reuse and the cost of mobility events as the main performance drivers. Our results show that mobility assumptions, event frequency, numerical solvers, and available hardware need to be considered together when developing metapopulation models for epidemic emergencies. In a synthesized setting with a disease outbreak in central Europe, we show how a worldwide pandemic setting with 12,865 patches and four billion state variables can be simulated in 24 seconds using GPU-based computing.
\end{abstract}



\end{frontmatter}

\section{Introduction}
\label{sec:introduction}

The International Disaster Database\footnote{https://www.emdat.be/} collects natural disasters based on the 2014 IRDR Peril Classification and Hazard Glossary~\cite{irdr_peril}, subdividing natural disasters into a hydrological, a meteorological, a geophysical, a biological, and a climatological family. Epidemics form a prominent example of biological disasters and the database has more than 1500 epidemics reported since 1900~\cite{tonnelier_what_2025}. In order to study the potential impact of epidemic developments on human societies, mathematical modeling has evolved as the principal tool for evidence generation~\cite{ecdc_npis_2024}. Yet, epidemic crises come with high time pressure on mathematical modelers, leaving little time to study scenarios and generate evidence on current and potential future developments. 

Over broad periods of the COVID-19 pandemic, disease dynamics have been largely heterogeneous across districts, regions, and countries.  And although large spatial heterogeneities have been observed, most models neglected a spatial dimension. A recent scoping review on COVID-19 models applied to Belgium found only three out of 37 compartmental-based models considering spatial resolution~\cite{bagaforo_scoping_2026}. Though, a relevant number of works have demonstrated that outbreak size estimates and reproduction numbers can be substantially biased if mobility is neglected~\cite{wesolowski2012quantifying,bajardi2011human,balcan2009multiscale,kraemer2020mobility,doi:10.1073/pnas.0510525103}. While the increased computational workload might have led to considerations that spatially resolved models were unsuitable for urgent epidemic settings, these considerations rather result from scripting-based single-core implementations and are not generally valid. To popularize spatially resolved models for future epidemics, we need highly efficient models that maximize the parallel performance on modern high performance computing infrastructure. To allow the study of spatially tailored disease mitigation strategies and enable timely parameter inference, the use of GPUs is particularly promising~\cite{Kovac2018}.

Metapopulation models are a particular class of spatially resolved models. They divide populations in a top-down approach into subpopulations that are spatially separated in \textit{patches}. Within these patches, the disease dynamics are modeled through patch-specific contact patterns. The patches are commonly connected through mobility terms, and this mobility can influence the timing and spatial development of epidemic waves~\cite{contreras2026mechanics,balcan2009multiscale}. As each patch represents a local subpopulation, the network of patches captures the spatial structure of disease spread.

The way mobility is represented is not merely an implementation choice as it can also change the simulated epidemic. Recurrent-trip assumptions, diffusive movement, and effective coupling formulations may lead to different outcomes~\cite{PhysRevX.1.011001,Citron2021}. Here, we separate \textit{implicit mobility} approaches, in which mobility-induced changes are modeled implicitly through patch-specific and network-based transition rates, from \textit{explicit mobility} approaches, in which mobile subpopulations are explicitly transferred between patches. For explicit mobility, we use the Lagrangian form, where each traveler remains always linked to its home patch. Our comparison therefore has to consider not only runtime but also possible differences in the dynamics produced by the mobility formulation.

Efficiency and performance of simulations can be improved both by the model formulation and by its (parallel) realization. Using a graph-based formulation with explicit, discrete mobility events, we recently suggested a stage-aligned Runge-Kutta scheme that leverages graph-specific structures to reduce computations and accelerate simulation~\cite{ZUNKER2027_stage_aligned}. Hybrid metapopulation approaches, on the other hand, can be considered as efficient multiscale approaches where different local model types are combined~\cite{bicker_hybrid_2024,kehrer_hybrid_2027}. Selecting an efficient programming language~\cite{pereira_energy_2017} and optimizing single-core performance can be considered a prerequisite of parallelization. Parallelization is naturally also already available with different epidemic model implementations. SimInf distributes stochastic metapopulation simulations with scheduled movement events across processors~\cite{JSSv091i12}, MEmilio uses intra- and inter-simulation parallelization for various models~\cite{bicker2026memilio}, GPU-based methods have been applied to age-structured epidemic ODE models and parameter fitting~\cite{Kovac2018}, and to simulate mesoscopic models in thousands of what-if scenarios~\cite{perumalla2021mesoscopic}.

Particular realizations of implicit and explicit mobility expose different parallel structures. Implicit formulations couple the patches at every right-hand-side evaluation~\cite{wang2018,liu_modelling_2022}. While most Lagrangian models of explicit mobility also require the coupled solution of subpopulations~\cite{ArinoVdD2003a,COSNER2009550,Arino2015,Rapaport2022}, we can decouple and parallelize population structures by approximating continuous mobility through discrete mobility events~\cite{ZUNKER2027_stage_aligned}. Using a graph-based approach, patch populations can evolve independently between events. Aligning the traveling subpopulation computation with the stages of the Runge-Kutta scheme applied to the patch-aggregated population, we can recover the exact state progression of travelers.

We introduce two metapopulation model structures with implicit and explicit mobility modeling in unified notation. Based on these structures, we provide a systematic analysis of how the mobility representations affect epidemic outcome, computational performance, memory usage, and parallel scalability. We develop and evaluate OpenMP and CUDA implementations for both cases and compare them with serial CPU execution. Both models have been implemented in C++ with OpenMP and CUDA in MEmilio~\cite{bicker2026memilio}. For implicit mobility, the patch-coupling matrix operations are parallelized. For explicit mobility, we use the independence between patch and traveler updates, that is obtained as a result of the discrete mobility exchange, and process several integration steps between mobility events together.

For our comparison, we use local SEIR dynamics with identical initial conditions on resident populations, mobility fractions, and numerical step sizes and analyze how much the mobility representation itself changes the trajectory. The performance study covers serial execution, OpenMP, and CUDA implementations. We provide a worldwide application with $12{,}865$ patches, weak and strong scaling on CPUs, subroutine timings, and roofline analyses for CPUs and GPUs.

The paper is structured as follows. In~\cref{sec:methods}, we introduce ODE-based metapopulation models using implicit and explicit mobility modeling. In~\cref{sec:complexity}, we conduct an extensive complexity analysis of the introduced metapopulation models, distinguishing two solution schemes for the explicit mobility model. In~\cref{sec:parallel-implementation}, we describe the parallel CPU and GPU implementations. In~\cref{sec:numerical-results}, we compare the epidemic outcomes of both mobility representations, show a worldwide application, and analyze runtime, parallel scalability, and roofline performance. In~\cref{sec:conclusion}, we summarize our findings.

\section{Methods}
\label{sec:methods}

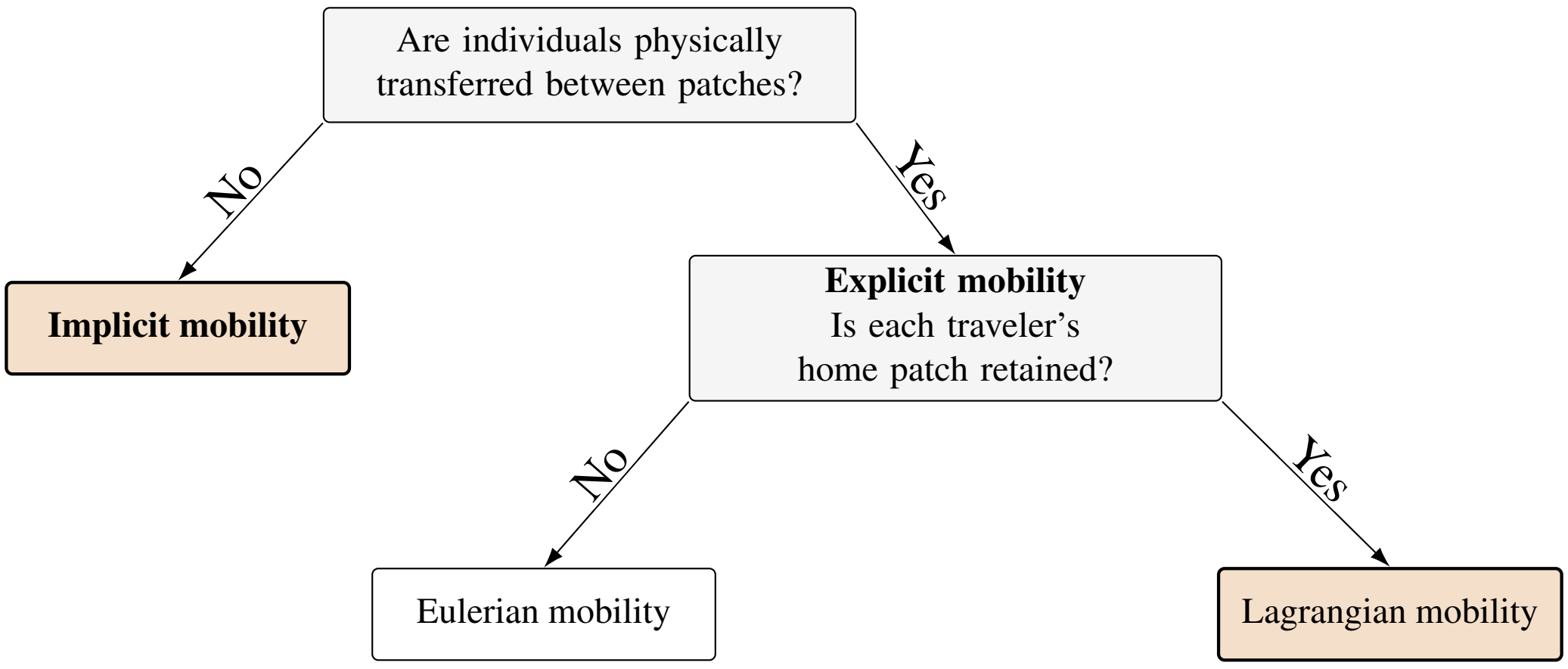
\begin{figure}[ht]
    \centering
    \begin{tikzpicture}[
        decision/.style={
            draw,
            rounded corners=1.5pt,
            align=center,
            text width=43mm,
            minimum height=10mm,
            inner xsep=5pt,
            inner ysep=3pt,
            fill=gray!8,
            font=\small
        },
        outcome/.style={
            draw,
            rounded corners=1.5pt,
            align=center,
            minimum width=30mm,
            minimum height=8mm,
            inner xsep=4pt,
            inner ysep=2pt,
            font=\small
        },
        studied/.style={
            outcome,
            fill=orange!30,
            line width=0.8pt
        },
        flow/.style={
            draw,
            -{Latex[length=1.8mm,width=1.2mm]}
        },
        answer/.style={
            fill=white,
            inner sep=1pt,
            font=\large
        }
    ]
        \node[decision] (transfer) at (0,0)
            {Are individuals physically transferred between patches?};
        \node[studied] (implicit) at (-3.6,-2.3)
            {\textbf{Implicit mobility}};
        \node[decision] (explicit) at (3.2,-2.3)
            {\textbf{Explicit mobility}\\Is each traveler's home patch retained?};
        \node[outcome] (eulerian) at (-0.4,-4.8)
            {Eulerian mobility};
        \node[studied] (lagrangian) at (7.0,-4.8)
            {Lagrangian mobility};

        \draw[flow] (transfer.south west)
            -- node[answer,sloped,above] {No} (implicit.north);
        \draw[flow] (transfer.south east)
            -- node[answer,sloped,above] {Yes} (explicit.north);
        \draw[flow] (explicit.south west)
            -- node[answer,sloped,above] {No} (eulerian.north);
        \draw[flow] (explicit.south east)
            -- node[answer,sloped,above] {Yes} (lagrangian.north);
    \end{tikzpicture}
    \caption{\textbf{Decision tree for mobility representations.} The highlighted outcomes, implicit mobility and explicit Lagrangian mobility, are considered in this work.}
    \label{fig:mobility-representations}
\end{figure}

Metapopulation models describe a spatially structured population as a collection of local subpopulations represented by the patches they are assigned to. Patches are connected through different ways of modeling mobility or their effects. More precisely, mobility enters these models either implicitly, by modeling mobility effects rather than mobility itself, or explicitly. In implicit modeling approaches, patches are coupled through their transition rates while the state variables of the patches remain associated with their home patch. In explicit approaches, subpopulations are explicitly moved between patches. Among explicit formulations one distinguishes Eulerian and Lagrangian settings. In the Eulerian case, a traveler joins the population of the destination patch upon executing an \textit{origin-destination} movement and then is no longer associated with its origin patch; in the Lagrangian case, the traveler keeps their origin or home-patch assignment even while moving to a new destination~\cite{ZUNKER2027_stage_aligned,Citron2021}. \cref{fig:mobility-representations} summarizes this classification and marks the two cases studied here, namely implicit mobility and explicit Lagrangian mobility. Let us highlight here that the terms \textit{implicit} and \textit{explicit} refer to the mobility representation, not the numerical solution scheme.

The local model used inside each patch is independent of this mobility distinction. The local dynamics can be written in many forms, including, its most common form of ODEs, but also stochastic differential equations, discrete-state stochastic processes such as continuous-time Markov chains, integro-differential equations, or hybrid formulations. In this paper, we restrict ourselves to deterministic ODE-based compartmental models. We further assume that all mobility formulations considered below are defined through the connections on the same directed graph
$\mathcal{G}=(\mathcal{P},\mathcal{E})$, with patch set
$\mathcal{P}=\{1,\ldots,N_P\}$, $N_P\in\mathbb{N}$, and edge set
$\mathcal{E}\subseteq\mathcal{P}\times\mathcal{P}$. We assume that $\mathcal{E}$ does not contain self-loops and define $\mathcal{E}_+=\{(p,p):p\in\mathcal{P}\}\cup\mathcal{E}$ as the set of edges of the graph with self-loops added. For each element $e\in\mathcal{E}_+$, we define a mobility coefficient $m_e\in\mathbb{R}_{\geq 0}$. We set $m_e=0$ for $e\in\mathcal{P}\times\mathcal{P}\setminus\mathcal{E}_+$. We denote $N_E=|\mathcal{E}|$. Note that the models presented here also allow for compartment (e.g., age or infection state) dependent mobility coefficients such that, e.g., older individuals commute less or severely infected individuals not at all. In this setting, the graph generalizes to a multigraph with generalized edge set $\mathcal{E}\subseteq\{(p,q;i)\in\mathcal{P}\times\mathcal{P}\times\mathbb{N}\}$; cf.~\cite{ZUNKER2027_stage_aligned}.

\subsection{ODE-based compartmental models in a single patch}
\label{sec:local-model}

\subsubsection*{A) General description}

As ODE-based metapopulation models are built from the individual contributions of compartmental models in different patches, we first consider a single patch population $N$ divided into $N_C\in\mathbb{N}$ compartments. Let $x_k(t)\in\mathbb{R}_{\geq0}$ denote the number of individuals in compartment or state $k\in\{1,\ldots,N_C\}$ at time $t$. The system state is
\begin{align}
    x(t)
    =\bigl(x_1(t),\ldots,x_{N_C}(t)\bigr)^T
    \in\mathbb{R}_{\geq0}^{N_C}.
    \label{eq:local-state}
\end{align}
In the following, we omit time arguments whenever they are unambiguous.

State changes are given by sign-free flow functions $f_{ij}:\mathbb{R}^{N_C}\times \mathbb{R}\rightarrow \mathbb{R}$ for a flow from state $i\in\{1,\ldots,N_C\}$ to state $j\in\{1,\ldots,N_C\}$. 
If no transition from state $i$ to state $j$ (equivalently considered as the tuple $(i,j)$) is possible, we set the corresponding flow function to $f_{ij}\equiv0$. We denote the number of nontrivial flow functions by $N_T\in\mathbb{N}$. For any compartment represented by index $k\in\{1,\ldots,N_C\}$, we index the set of (nontrivial) incoming flows through their prior state by $\mathcal{I}_k\subset\{1,\ldots,N_C\}$ and the set of (nontrivial) outgoing flows through their next state by $\mathcal{O}_k\subset\{1,\ldots,N_C\}$. The change is then obtained as
\begin{align}
\label{eq:generic_ode_model}
\begin{aligned}
    \frac{\d x_k(t)}{\d t} = \sum_{j \in \mathcal{I}_k} f_{jk}\left(x(t), t\right) - \sum_{l \in \mathcal{O}_k} f_{kl}\left(x(t), t\right).
\end{aligned}
\end{align}

Assuming that each flow $f_{ij}$ has a prescribed source and target, it appears twice in the system described by~\cref{eq:generic_ode_model}; once with a negative sign (for $x_i$) and once with a positive sign (for $x_j$). Then, the population is preserved by construction, i.e.,
\begin{align}\label{eq:preserv}
    \sum_{k=1}^{N_C}\frac{\d x_k(t)}{\d t}=0.
\end{align}

\begin{remark}
Note that~\cref{eq:preserv,eq:generic_ode_model} mean that there is no in- or outflow into or from the system. In order to realize those as, e.g., births, immigration, or natural deaths, auxiliary states can be included in~\cref{eq:local-state} to preserve the population in the system sense. While the system's total population remains constant, the sum of non-auxiliary compartments in the patch may change over time.
\end{remark}

Defining $\Gamma:\{1,\ldots,N_C\}\times \{1,\ldots,N_C\}\rightarrow \{1,\ldots,N_T\}$ as the function mapping all tuples $(i,j)$ with $f_{ij}\not\equiv 0$ onto an ordered list, we can index the flow functions by a single index $r_{(i,j)}=\Gamma(i,j)\in\{1,\ldots,N_T\}$, referring to its position in the ordered list of tuples. We organize the coefficients of the directed flows from~\cref{eq:generic_ode_model} in $B\in\mathbb{R}^{N_C\times N_T}$, where $B_{ij}\in\{-1,0,1\}$, $i\in\{1,\ldots,N_C\}$, $j\in\{1,\ldots,N_T\}$. Each row of $B$ then contains the (nontrivial) coefficients of incoming and outgoing flows and each column determines how the compartment populations change when one individual moves along the corresponding transition. More precisely, for each transition $r_{(i,j)}\in\{1,\ldots,N_T\}$, $i\in\{1,\ldots,N_C\}$ denotes its source-compartment index and $j\in\{1,\ldots,N_C\}$ its target-compartment index. For better readability, we drop the double index on $r$. Then, we have $B_{ir}=-1$, $B_{jr}=1$, and all other entries in column $r$ are zero. 

The common assumption that any flow is proportional to its source compartment and that its intensity acts homogeneously on all individuals of that same compartment means that we can write
\begin{align}
    f_r(x(t),t)
    =x_{i}(t)d_r(x(t),t),
    \qquad r=1,\ldots,N_T;
    \label{eq:local-flow}
\end{align}
also cf.~\cite[Assumption~3.1]{ZUNKER2027_stage_aligned}. Here, $d_r(x(t),t)\geq0$ is the per-capita transition intensity. 

The flow functions describing the transitions in the population can be classified as either interaction-independent first or interaction-dependent second order transitions. With the notation of~\cref{eq:local-flow}, interaction-independent first-order (linear) transitions are of the form
\begin{align}\label{eq:1st_adopt}
f_r(x(t),t)=x_i(t)d_r(t),
\end{align}
i.e., $d_r(t)$ only depends on the time and the flow function effectively only on the current source compartment. Interaction-dependent second order transitions are generally given by
\begin{align}\label{eq:2nd_adopt}
f_r(x(t),t)=x_i(t)d_r(x(t),t)=x_i(t)\sum_{l\in\Psi_{ij}}c_{ij,l}(t)x_l(t),
\end{align}
where $\Psi_{ij}\subset\{1,\ldots,{N_C}\} \setminus \{i\}$ contains the indices of the compartments that may influence the transition from state $x_i$ to~$x_j$. The coefficients $c_{ij,l}(t):\mathbb{R}\rightarrow \mathbb{R}_{\geq 0}$ may account for contact rates and scaling factors.

Defining $D(x(t),t)\in\mathbb{R}^{N_T\times N_C}$ by
\begin{align}
    D_{rs}(x(t),t)
    =
    \begin{cases}
        d_r(x(t),t), & s=i,\\
        0, & \text{otherwise},
    \end{cases}
    \label{eq:local-rate-matrix}
\end{align}
and $f(x(t),t)=(f_1(x(t),t),\ldots,f_{N_T}(x(t),t))^T$, the local dynamics are
\begin{align}
    \frac{\mathrm{d}x(t)}{\mathrm{d}t}
    =Bf(x(t),t)
    =BD(x(t),t)x(t).
    \label{eq:local-model}
\end{align}
We assume that each $d_r$ is continuous in $t$ and globally Lipschitz continuous in $x$, such that~\cref{eq:local-model} admits a unique solution.

\subsubsection*{B) The age-stratified SEIR model}
    To illustrate our notation, consider an age-stratified SEIR model with $N_G$ age groups. The total number of compartments is $N_C=4N_G$ and we write 
\begin{align*}
    x(t) = (S_1(t), E_1(t), I_1(t), R_1(t), \ldots, S_{N_G}(t), E_{N_G}(t), I_{N_G}(t), R_{N_G}(t))^T.
\end{align*}
Let $N_i=S_i+E_i+I_i+R_i\geq0$ be the population of age group $i$, $\rho_i \in [0,1]$ its infection risk per contact and $\phi_{ij}(t)\in \mathbb{R}_{\ge0}$ its contact rate with age group $j$. With $T_{E_i}{>0}$ and $T_{I_i}{>0}$ we denote the mean durations of the exposed and infectious states, respectively. The force of infection is 
\begin{align}
    \lambda_i(x(t),t)= \rho_i\sum_{j=1}^{N_G}\phi_{ij}(t)\frac{I_j(t)}{N_j(t)},
    \label{eq:local-foi}
\end{align}
where the fraction is set to zero if $N_j=0$. Individuals cannot change their age group, so for~\cref{eq:generic_ode_model}, we have
\begin{align*}
f_{4i-3, 4i-2}(x(t),t)=\lambda_i(x(t),t)S_i(t),\quad f_{4i-2, 4i-1}(x(t),t)=T_{E_i}^{-1}{E_i(t)},\quad f_{4i-1, 4i}(x(t),t)=T_{I_i}^{-1}{I_i(t)},\quad i=1,\ldots, N_G,
\end{align*}
and for~\cref{eq:local-model}, we have
\begin{align}
\label{eq:local-rate-matrix-seir}
    B=\begin{pmatrix}
        B_1 &&\\
        &\ddots&\\
        &&B_{N_G}
    \end{pmatrix},\quad
    B_i=\begin{pmatrix}
        -1 &0&0 \\
        1 &-1 &0 \\
        0&1& -1 \\
        0&0& 1
    \end{pmatrix},  \quad 
    D=\begin{pmatrix}
        D_1 && \\
        & \ddots & \\
        &&D_{N_G}
    \end{pmatrix}, \quad
    D_i = \begin{pmatrix}
        \lambda_i(x(t),t)& 0 & 0 &0\\
        0 & T_{E_i}^{-1} & 0 &0\\
        0 & 0 & T_{I_i}^{-1} & 0 
    \end{pmatrix}.
\end{align}
The ODE system reads
\begin{align}
\begin{aligned}
    \frac{\d S_i(t)}{\d t} &= -\lambda_i(x(t),t)S_i(t),\\
        \frac{\d E_i(t)}{\d t} &= \lambda_i(x(t),t)S_i(t)-\frac{1}{T_{E_i}}E_i(t),\\
        \frac{\d I_i(t)}{\d t} &= \frac{1}{T_{E_i}}E_i(t)-\frac{1}{T_{I_i}}I_i(t),\\
        \frac{\d R_i(t)}{\d t} &= \frac{1}{T_{I_i}}I_i(t),
        \qquad i=1,\ldots,N_G.
        \label{eq:local-seir}
        \end{aligned}
\end{align}

\subsection{ODE-based metapopulation models}
\label{sec:metapopulation-models}

Metapopulation models are built from the local patch contributions and the global system state, with individuals assigned to their origin (or home) patch, is
\begin{align}\label{eq:metapop_1n}
x(t)=(x^{(1)T}(t),\ldots,x^{(N_P)T}(t))^T,
\end{align}
where each $x^{(p)}(t)\in\mathbb{R}^{N_C}_{\geq 0}$, $p\in\{1,\ldots,N_P\}$, takes the form of~\cref{eq:local-state}.

For all metapopulation models presented here, we assume that every patch uses the same compartmental and transition structure. For $p\in\mathcal{P}$, we denote the locally parametrized transition intensities of the form of~\cref{eq:local-rate-matrix} by $d_r^{(p)}$ and $D^{(p)}$.

\subsubsection{Implicit mobility}
\label{sec:implicit-mobility}

\subsubsection*{A) General description}

The implicit mobility formulation builds initially on the metapopulation SIR model of Wang et al.~\cite{wang2018} and the related area-based SEIR model of Liu et al.~\cite{liu_modelling_2022}, with modifications made by Gerstein~\cite{gerstein_MA}. Our formulation considers origin and destination phases and evaluates the transition intensities based on the populations weighted proportionally with the different phases.

\begin{remark}
Note that with mass action incidence and a constant $\alpha$-term as presented in~\cite{wang2018,liu_modelling_2022}, we risk missing the correct normalization factor; cf. the discussion in~\cite[Sec.~3.1.1--3.1.3]{gerstein_MA}. As a result of an incorrect normalization, the reproduction number erroneously scales with the spatial resolution and the number of patches; see~\cite[Table~5.2]{gerstein_MA}.
\end{remark}

The state vector of the implicit mobility formulation is of the form of~\cref{eq:metapop_1n}. We assume that mobility exchange is described by a constant matrix $M\in[0,1]^{N_P\times N_P}$, where the entry $m_{pq}$ corresponds to the fraction of residents of Patch~$p$ ``present'' in Patch~$q$ during the mobility phase. Note that the presence is not enforced explicitly and that its effect is modeled implicitly through proportional contact intensities, therefore coining the corresponding approaches as models with implicit mobility (effects). Note that $m_{pq}=0$ if $(p,q)\notin\mathcal{E}$ and $p\neq q$. We require
\begin{align}
    \sum_{\substack{q=1\\q\neq p}}^{N_P}m_{pq}  \leq 1,  \qquad m_{pp} =
    1-\sum_{\substack{q=1\\q\neq p}}^{N_P}m_{pq},    \qquad p\in\mathcal{P}.
    \label{eq:implicit-mobility-fractions}
\end{align}
Thus, the entries in each row of $M$ sum to one. The rows of the mobility matrix $M$ contain for each patch the fraction of non-traveling individuals $m_{pp}$ as well as all outgoing connections $m_{pq}$. Thus, $M$ contains $N_E+N_P$ nontrivial values and the share of nontrivial values is $(N_E+N_P)/N_P^2$.

For $p,q\in\mathcal{P}$, we define the ``virtual'' populations
\begin{align}
    x_{\mathrm{impl}}^{(p;q)}(t)= m_{pq}x^{(p)}(t)
    \label{eq:implicit-virtual-travelers}
\end{align}
from origin $p$, acting at destination $q$. 

The population ``virtually''  present in Patch~$q$ is
\begin{align}
    \widetilde{x}_{\mathrm{impl}}^{(q)}(t)
    =
    \sum_{p=1}^{N_P}x_{\mathrm{impl}}^{(p;q)}(t)
    =
    \sum_{p=1}^{N_P}m_{pq}x^{(p)}(t).
    \label{eq:implicit-present-population}
\end{align}

In matrix notation, we collect the resident patch states $x^{(p)}(t)$ from~\cref{eq:metapop_1n} as the rows of $X(t)\in\mathbb{R}_{\geq0}^{N_P\times N_C}$, i.e.,
\begin{align}
    X(t)=\begin{pmatrix}
        x^{(1)\,T}\\
        \vdots\\
        x^{(N_P)\,T}
    \end{pmatrix}.
\end{align}
Likewise, we collect the populations ``virtually'' present in the patches from~\cref{eq:implicit-present-population} as the rows of $\widetilde{X}\in\mathbb{R}_{\geq0}^{N_P\times N_C}$ by
\begin{align}
    \widetilde{X}
    &=
    M^T X.
    \label{eq:implicit-patch-aggregation}
\end{align}

\begin{remark}\label{rem:mobcoeff}
In the case of compartment-specific fractions $m_{pq}\in[0,1]^{N_C}$, our graph generalizes to a multigraph with up to $N_C$ edges per node pair and \cref{eq:implicit-virtual-travelers} to $x_{\mathrm{impl}}^{(p;q)}(t)=m_{pq}\odot x^{(p)}(t)$, where $\odot$ represents element-wise multiplication. In addition, the product $M^TX$ needs to be replaced by one product per compartment.
\end{remark}

Motivated by round trip-based mobility, we assume for $\theta\in[0,1]$ that (virtual) presence only drives transitions on a fraction $(1-\theta)$ of unit time and that for the remaining fraction $\theta$, transitions are solely driven by the original population, i.e., we define local transition intensity matrices dependent on global $x(t)$ as
\begin{align}
    \overline{D}_{\mathrm{impl}}^{(p)}(x(t),t)
    =
    \theta
    D^{(p)}(x^{(p)}(t),t)
    +(1-\theta)\sum_{q=1}^{N_P}    m_{pq}D^{(q)}
    \left(\widetilde{x}_{\mathrm{impl}}^{(q)},t\right).
    \label{eq:implicit-intensity-theta}
\end{align}
Note that the scaling term in the sum is an already integrated weighting for $x^{(p)}$ if the intensity is multiplied inside the flow; cf.~\cref{eq:local-flow}. The sum includes for $q=p$ the acting intensity on individuals staying in its original patch and for $q\neq p$ the intensity acting on out-commuters virtually present in Patch~$q$.

Splitting the local transition intensity matrix according to first and second order transitions as in~\cref{eq:1st_adopt,eq:2nd_adopt} and assuming that first-order, interaction-independent intensities, such as progression or recovery, are equal for all patches and independent of $t$, we obtain with~\cref{eq:implicit-mobility-fractions} that
\begin{align}
\overline{D}_{\mathrm{impl}}^{(p)}(x(t),t)
    =D_{\mathrm{ind}}
    + \theta
    D_{\mathrm{dep}}^{(p)}(x^{(p)}(t),t)
    +(1-\theta)\sum_{q=1}^{N_P}    m_{pq}D_{\mathrm{dep}}^{(q)}
    \left(\widetilde{x}_{\mathrm{impl}}^{(q)},t\right).
    \label{eq:implicit-intensity-decomposition}
\end{align}
Here, $D_{\mathrm{ind}}$ contains interaction-independent transitions whereas $D_{\mathrm{dep}}^{(q)}$ contains intensities dependent on the population and determined by the local conditions in Patch~$q$. Due to the definition (see~\cref{eq:1st_adopt,eq:2nd_adopt}), a transition is either interaction-independent or interaction-dependent. Let $\mathcal{R}=\{1,\ldots,N_T\}$ denote the set of transition indices.
Therefore, we can write $\mathcal{R} = \mathcal{R}_{\mathrm{ind}} \cup \mathcal{R}_{\mathrm{dep}}$ with $\mathcal{R}_{\mathrm{ind}} \cap \mathcal{R}_{\mathrm{dep}}=\emptyset$ and define 
\begin{align}
    d_{\mathrm{ind},r}=
    \begin{cases}
        d_r^{(q)}, & r\in\mathcal{R}_{\mathrm{ind}},\\
        0, & r\in\mathcal{R}_{\mathrm{dep}},
    \end{cases}
    \qquad
    d_{\mathrm{dep},r}^{(q)}(x,t)=
    \begin{cases}
        0, & r\in\mathcal{R}_{\mathrm{ind}},\\
        d_r^{(q)}(x,t), & r\in\mathcal{R}_{\mathrm{dep}},
    \end{cases}
    \label{eq:implicit-intensity-split}
\end{align}
such that $D^{(q)}=D_{\mathrm{ind}}+D_{\mathrm{dep}}^{(q)}$.

The progression of states in the implicit mobility effects model then is defined by a system of dimension $N_PN_C$ with, for $p\in\mathcal{P}$,
\begin{align}
    \frac{\mathrm{d}x^{(p)}(t)}{\mathrm{d}t} =
    B\overline{D}_{\mathrm{impl}}^{(p)}(x(t),t)x^{(p)}(t).
    \label{eq:implicit-model}
\end{align}

\subsubsection*{B) Computational aspects}

The matrices $D^{(q)}$ are sparse with a fraction of $N_C^{-1}$ nonzero values. For computations with dense structures and our complexity analysis, we define $\widehat{\Lambda}^{\mathrm{home}},\widehat{\Lambda}^{\mathrm{mob}} \in\mathbb{R}_{\geq0}^{N_P\times N_T}$ with entries
\begin{align}
\widehat{\Lambda}_{qr}^{\mathrm{home}}(x(t),t)
    =
    d_{\mathrm{dep},r}^{(q)}\left(x^{(q)}(t),t\right)
    \quad\textnormal{and}\quad
    \widehat{\Lambda}_{qr}^{\mathrm{mob}}(x(t),t)
    =
    d_{\mathrm{dep},r}^{(q)}\left(\widetilde{x}_{\mathrm{impl}}^{(q)}(t),t\right),
    \qquad q\in\mathcal{P},\quad r=1,\ldots,N_T.
    \label{eq:implicit-patch-intensities}
\end{align}
For the home contributions, we have a direct redefinition while for the mobility contributions these patch-specific intensities need to be mapped back to the residents' home patches, i.e.,
\begin{align}
{\Lambda}^{\mathrm{home}}(x(t),t)
=\widehat{\Lambda}^{\mathrm{home}}(x(t),t)
\quad\textnormal{and}\quad
    \Lambda^{\mathrm{mob}}(x(t),t)
    =
    M\widehat{\Lambda}^{\mathrm{mob}}(x(t),t).
    \label{eq:implicit-intensity-aggregation}
\end{align}
For residents of Patch~$p$ and transition $r$, this gives $\Lambda^{\mathrm{mob}}_{pr}=\sum_{q=1}^{N_P}m_{pq}d_{\mathrm{dep},r}^{(q)}(\widetilde{x}_{\mathrm{impl}}^{(q)},t)$. This is the mobility-weighted average intensity acting on the residents of Patch~$p$ before multiplication by the phase weight $1-\theta$. The complete effective per-capita transition intensities $\overline{d}_{\mathrm{impl},r}^{(p)}$ are available row-wise in each Patch~$p\in\mathcal{P}$ with
\begin{align}
    \overline{\Lambda}(x(t),t)
    &=
    \mathbf{1}_{N_P} d_{\mathrm{ind}}^T
    +\theta \Lambda^{\mathrm{home}}(x(t),t)
    +(1-\theta)\Lambda^{\mathrm{mob}}(x(t), t).
    \label{eq:implicit-effective-intensity}
\end{align}
Here, $\mathbf{1}_{N_P}\in\mathbb{R}^{N_P}$ denotes the vector of ones, $d_{\mathrm{ind}}=(d_{\mathrm{ind},1},\ldots,d_{\mathrm{ind},N_T})^T$.

Due to the structure of the flows (see~\cref{eq:local-flow}), row~$p$ of $\overline{\Lambda}$ directly yields the flows $f_r^{(p)}=x_i^{(p)}\,\overline{\Lambda}_{pr}$, such that the right-hand side of~\cref{eq:implicit-model} is evaluated as $Bf^{(p)}$ in $\mathcal{O}(N_C+N_T)$ operations. Instead of explicitly building the matrix $\overline{D}_{\mathrm{impl}}^{(p)}$, we only store the columns of $\overline{\Lambda}$ with indices $\mathcal{R}_{\mathrm{dep}}$ since all columns $r\in\mathcal{R}_{\mathrm{ind}}$ of $\widehat{\Lambda}^{\mathrm{home}}$, $\widehat{\Lambda}^{\mathrm{mob}}$, $\Lambda^{\mathrm{home}}$, and $\Lambda^{\mathrm{mob}}$ are zero. 

\subsubsection*{C) The age-stratified SEIR metapopulation model with implicit mobility effects}
Let us consider local SEIR dynamics as in~\cref{sec:local-model}B.
On the metapopulation level, we then have 
\begin{align*}
    X(t)=\begin{pmatrix}
        S_1^{(1)}(t) & E_1^{(1)}(t) & I_1^{(1)}(t) & R_1^{(1)}(t) & \cdots & S^{(1)}_{N_G}(t) & E^{(1)}_{N_G}(t) & I^{(1)}_{N_G}(t) & R^{(1)}_{N_G}(t)\\
        \vdots\\
        S_1^{(N_P)}(t) & E_1^{(N_P)}(t) & I_1^{(N_P)}(t) & R_1^{(N_P)}(t) &  \cdots & S^{(N_P)}_{N_G}(t) & E^{(N_P)}_{N_G}(t) & I^{(N_P)}_{N_G}(t) & R^{(N_P)}_{N_G}(t) 
    \end{pmatrix}
\end{align*}
with $N^{(q)}_{j}=S_j^{(q)}(t)+E_j^{(q)}(t)+I_j^{(q)}(t)+R_j^{(q)}(t)$ and 
\begin{align*}
    \widetilde{X}(t)=M^TX(t)=\begin{pmatrix}
        \sum_{p=1}^{N_P}m_{p1}S_1^{(p)}(t) & \sum_{p=1}^{N_P}m_{p1}E_1^{(p)}(t) & \cdots & \sum_{p=1}^{N_P}m_{p1}R_{N_G}^{(p)}(t)\\
        \vdots\\
        \sum_{p=1}^{N_P}m_{pN_P}S_1^{(p)}(t) & \sum_{p=1}^{N_P}m_{pN_P}E_1^{(p)}(t) & \cdots & \sum_{p=1}^{N_P}m_{pN_P}R_{N_G}^{(p)}(t)    
    \end{pmatrix}
\end{align*}
with $\widetilde{N}^{(q)}_{j}=\sum_{p=1}^{N_P}m_{pq}N_j^{(p)}$. Given that the interaction-dependent transition intensities in the SEIR model depend on the infectious individuals, we define the submatrices
\begin{align*}
    X_I(t)=\begin{pmatrix}
        I_1^{(1)}(t) & \cdots & I^{(1)}_{N_G}(t)\\
        \vdots\\
        I_1^{(N_P)}(t) & \cdots & I^{(N_P)}_{N_G}(t) 
    \end{pmatrix}\quad\textnormal{and}\quad
    \widetilde{X}_I(t)=M^TX_I(t)=\begin{pmatrix}
        \sum_{p=1}^{N_P}m_{p1}I_1^{(p)}(t) & \cdots & \sum_{p=1}^{N_P}m_{p1}I_{N_G}^{(p)}(t)\\
        \vdots\\
        \sum_{p=1}^{N_P}m_{pN_P}I_1^{(p)}(t)& \cdots & \sum_{p=1}^{N_P}m_{pN_P}I_{N_G}^{(p)}(t) 
    \end{pmatrix}.
\end{align*}
The infectious fraction of age group $j$ among the population virtually present in Patch~$q$ is 
\begin{align}\label{eq:Uqj}
    U_{qj}(t) = \frac{\widetilde{X}_{I,qj}(t)}{\widetilde{N}^{(q)}_{j}},
\end{align}
with $U_{qj}=0$ if $\widetilde{N}^{(q)}_{j}=0$.
We now define the patch-independent transition indices. For each age group $i\in\{1,\ldots,N_G\}$, let $r^{SE}_i=\Gamma(4i-3,4i-2)$, $r_i^{EI}=\Gamma(4i-2,4i-1)$, and $r_i^{IR}=\Gamma(4i-1,4i)$ denote the indices of the transitions $S_i\rightarrow E_i$, $E_i\rightarrow I_i$, and $I_i\rightarrow R_i$, respectively. The last two are interaction-independent, i.e., $\mathcal{R}_{\mathrm{dep}}=\{r^{SE}_1,\ldots,r^{SE}_{N_G}\}$ and $\mathcal{R}_{\mathrm{ind}}=\{r_i^{EI},r_i^{IR}: i=1,\ldots,N_G\}$. For transition $r\in\mathcal{R}_{\mathrm{dep}}$ with age group $i\in\{1,\ldots,N_G\}$, we have 
\begin{align*}
    \widehat{\Lambda}^{\mathrm{mob}}_{qr}=\rho_i^{(q)}\sum_{j=1}^{N_G} \phi_{ij}^{(q)}(t)U_{qj}(t).
\end{align*}
We assume that $\rho_i^{(q)}=\rho_i$ and $\phi_{ij}^{(q)}(t)=\phi_{ij}$ are patch- and time-independent. Thus, the sum over $q$ in~\cref{eq:implicit-intensity-aggregation} can be moved inside, and 
\begin{align}
    \Lambda^{\mathrm{mob}}_{pr^{SE}_i}(t) =\sum_{q=1}^{N_P}m_{pq}\widehat{\Lambda}^{\mathrm{mob}}_{qr^{SE}_i}(t)=\rho_i\sum_{j=1}^{N_G} \phi_{ij}V_{pj}(t) \quad\textnormal{with}\quad V_{pj}(t) = \sum_{q=1}^{N_P}m_{pq}U_{qj}(t),
    \label{eq:implicit-seir-V}
\end{align}
that is, $V(t)=MU(t)$.
We have $\Lambda^{\mathrm{home}}_{pr^{SE}_i}=\rho_i\sum_j\phi_{ij}I^{(p)}_{j}(t)/N^{(p)}_{j}$. Inserting both terms and $d_{\mathrm{ind}}$ into~\cref{eq:implicit-effective-intensity} yields, for residents of Patch~$p$,
\begin{align}
    \overline{\Lambda}_{pr^{SE}_i}(x(t),t)&=\lambda_i^{(p)}(x(t),t)=\rho_i\sum_{j=1}^{N_G}\phi_{ij}\left(\theta\frac{I^{(p)}_{j}(t)}{N^{(p)}_{j}}+(1-\theta)V_{pj}(t)\right),
    \quad    \overline{\Lambda}_{pr_i^{EI}}&=d_{\mathrm{ind},r_i^{EI}}=T_{E_i}^{-1},
    \quad    \overline{\Lambda}_{pr_i^{IR}}&=d_{\mathrm{ind},r_i^{IR}}=T_{I_i}^{-1}.
    \label{eq:implicit-seir-force}
\end{align}
Here, $M$ is effectively multiplied with the infectious fractions $U$ instead of a direct application to $\widehat{\Lambda}^{\mathrm{mob}}$. As a consequence, the home and mobility contributions only need a single contact sum per patch.
The full dynamics are then given with~\cref{eq:local-seir} for $p\in\mathcal{P}$ with $\lambda_i$ replaced by $\lambda_i^{(p)}$.

\subsubsection{Explicit Lagrangian mobility}
\label{sec:explicit-mobility}

\subsubsection*{A) General description}

The explicit Lagrangian formulation tracks both the home patch affiliation and current location. 
For $(p,q)\in\mathcal{E}_+$, the state $x^{(p;q)}\in\mathbb{R}_{\geq0}^{N_C}$ represents residents of $p$ currently located in $q$. In particular, $x^{(p;p)}$ represents residents of $p$ who are currently in their home patch.
The total state of residents $x^{(p)}$ of $p$ and the patch-aggregated state $\widetilde{x}^{(q)}$ of the population present in $q$ are
\begin{align}
    x^{(p)}(t)
    =\sum_{q:(p,q)\in\mathcal{E}_+}x^{(p;q)}(t)
    \qquad\textnormal{and}\qquad
    \widetilde{x}^{(q)}(t)
    =\sum_{p:(p,q)\in\mathcal{E}_+}x^{(p;q)}(t)
    \label{eq:explicit-populations}
\end{align}
respectively. Due to the homogeneous-mixing (or \textit{compartment-level homogeneity} as formulated in~\cite[Assumption~3.1]{ZUNKER2027_stage_aligned}) assumption, all origin-specific subpopulations $x^{(p;q)}$ present in Patch~$q$ are subject to the same per-capita transition intensity matrix $D^{(q)}(\widetilde{x}^{(q)}(t),t)$.

Following Zunker et al.~\cite{ZUNKER2027_stage_aligned}, we assume that mobility occurs instantaneously at the times of a sequence $(\tau_a)_{a\in\mathbb{N}_0}$ with $0=\tau_0<\tau_1<\tau_2<\cdots$. Between two consecutive events, the origin-specific states evolve continuously under the local ODEs. For simplicity, we assume that each $\tau_a$, $a\in\mathbb{N}_0$, either represents a departure or a return event for all patches. Thus, on $I_a=[\tau_a,\tau_{a+1})$, $x^{(p;q)}$ from~\cref{eq:explicit-populations} evolves with
\begin{align}
    \frac{\d x^{(p;q)}(t)}{\d t}
    &=
    BD^{(q)}\left(\widetilde{x}^{(q)}(t),t\right)x^{(p;q)}(t),  \qquad (p,q)\in\mathcal{E}_+
    \label{eq:explicit-model-standard}
\end{align}
and $\widetilde{x}^{(q)}$ from~\cref{eq:explicit-populations} evolves with
\begin{align}
    \frac{\d \widetilde{x}^{(q)}(t)}{\d t}
    &=
    BD^{(q)}\left(\widetilde{x}^{(q)}(t),t\right)
      \widetilde{x}^{(q)}(t), \qquad q\in\mathcal{P},
    \label{eq:explicit-model-zunker}
\end{align}
given new initial conditions for each $\tau_a$, $a\in\mathbb{N}_0$. The second equation follows by summing~\cref{eq:explicit-model-standard} over all origins. 

For $a\geq1$, the superscript $-$ denotes the pre-event limit $x^{(p;q)}(\tau_a^-)=\lim_{t\uparrow\tau_a}x^{(p;q)}(t)$, whereas $+$ denotes the post-event limit $x^{(p;q)}(\tau_a^+)=\lim_{t\downarrow\tau_a}x^{(p;q)}(t)$. At the initial time $\tau_0=0$, $x^{(p;q)}(\tau_0^-)$ denotes the initial state before a possible mobility update and is not a left-sided limit of the simulated trajectory. We use the convention $x^{(p;q)}(\tau_a)=x^{(p;q)}(\tau_a^+)$, so each mobility event is effective at its scheduled time. In particular, if the final simulation time $t_{\max}\in\mathbb{R}$ coincides with an event time $\tau_a$, the reported final state includes the final mobility update at $\tau_a$.

For every $(p,q)\in\mathcal{E}$, let
\begin{align}
    m_a^{(p;p\to q)},\,
    m_a^{(p;q\to p)}
    \in[0,1]
    \label{eq:explicit-event-fractions}
\end{align}
denote the departure and return fractions at event time $\tau_a$. We assume that the fractions are independent of the compartment. For departure events $a$, we use $m_a^{(p;p\to q)}=m_{pq}$ with the fractions $m_{pq}$ from~\cref{eq:implicit-mobility-fractions} and $m_a^{(p;q\to p)}=0$. The return events move all travelers back to their home patch, i.e., $m_a^{(p;p\to q)}=0$ and $m_a^{(p;q\to p)}=1$ for $a$ representing a return event. For the generalization with compartment-specific fractions, see~\cite[Eq.~(10)]{ZUNKER2027_stage_aligned}. We define the net change of the traveler state by
\begin{align}
    \Delta_a^{(p;q)} = 
    m_a^{(p;p\to q)}  x^{(p;p)}(\tau_a^-) 
    - m_a^{(p;q\to p)} x^{(p;q)}(\tau_a^-).
    \label{eq:explicit-event-delta}
\end{align}
Therefore, the event update is given by
\begin{align}
\begin{aligned}
    x^{(p;q)}(\tau_a^+)
    &=
    x^{(p;q)}(\tau_a^-)  + \Delta_a^{(p;q)}, \qquad\qquad q:(p,q)\in\mathcal{E},\\
    x^{(p;p)}(\tau_a^+)
    &=
    x^{(p;p)}(\tau_a^-) -\hspace*{-0.2cm}\sum_{q:(p,q)\in\mathcal{E}} \hspace*{-0.2cm}\Delta_a^{(p;q)}.
    \end{aligned}
    \label{eq:explicit-event-update}
\end{align}

The update conserves the resident population of every home patch,
\begin{align}
    \sum_{q:(p,q)\in\mathcal{E}_+}x^{(p;q)}(\tau_a^+)
    =
    \sum_{q:(p,q)\in\mathcal{E}_+}x^{(p;q)}(\tau_a^-),
    \qquad p\in\mathcal{P}.
    \label{eq:explicit-event-conservation}
\end{align}
After the update, the patch aggregates are recomputed using~\cref{eq:explicit-populations}. Unlike the ``virtual'' populations in the implicit setup (see~\cref{eq:implicit-virtual-travelers}), the traveler populations are retained as independent state variables and advanced separately.

Defining $x^{(p;q)}=0$ if $(p,q)\notin\mathcal{E}_+$, we arrange the origin-destination states $x^{(p;q)}\in\mathbb{R}_{\geq0}^{N_C}$ as the entries of the matrix
\begin{align}
    \mathcal{X}(t)=\begin{pNiceMatrix}[first-row,first-col,last-row,last-col,cell-space-limits=2pt]
    \CodeBefore
      \cellcolor{black!10}{1-1,2-2,4-4}
    \Body
      & \scriptstyle q=1 & \scriptstyle q=2 & & \scriptstyle q=N_P & \\
    \scriptstyle p=1 & x^{(1;1)}(t) & x^{(1;2)}(t) & \cdots & x^{(1;N_P)}(t) & \;\rightarrow x^{(1)}(t)\\
    \scriptstyle p=2 & x^{(2;1)}(t) & x^{(2;2)}(t) & \cdots & x^{(2;N_P)}(t) & \;\rightarrow x^{(2)}(t)\\
      & \vdots & \vdots & \ddots & \vdots & \;\;\vdots\\
    \scriptstyle p=N_P & x^{(N_P;1)}(t) & x^{(N_P;2)}(t) & \cdots & x^{(N_P;N_P)}(t) & \;\rightarrow x^{(N_P)}(t)\\
      & \downarrow & \downarrow & & \downarrow & \\[-2pt]
      & \tilde x^{(1)}(t) & \tilde x^{(2)}(t) & & \tilde x^{(N_P)}(t) &
  \end{pNiceMatrix}
  \label{eq:explicit-od-matrix}
\end{align}
Row $p$ contains the residents of Patch~$p$ at all locations, and column $q$ the population present in $q$. By~\cref{eq:explicit-populations} the row sums are the resident states $x^{(p)}$ and the column sums are the aggregates $\widetilde{x}^{(q)}$. The shaded diagonal holds the residents at home, and the off-diagonal entries are the travelers.
Mobility events transfer populations within a row. By~\cref{eq:explicit-event-delta} and \cref{eq:explicit-event-update}, the departure fraction $m_a^{(p;p\rightarrow q)}$ moves part of the diagonal entry $x^{(p;p)}$ to the off-diagonal entry $x^{(p;q)}$ of the same row, and the return fraction $m_a^{(p;q\rightarrow p)}$ moves part of $x^{(p;q)}$ back to $x^{(p;p)}$.

\subsubsection*{B) Computational aspects}

The explicit Lagrangian system defined by~\cref{eq:explicit-model-standard,eq:explicit-event-update} might be solved by direct global integration of all origin-destination states $x^{(p;q)}$, $(p,q)\in\mathcal{E}_+$ with an explicit Runge-Kutta (RK) method acting on a $(N_P+N_E)N_C$-dimensional state. At each RK stage, the integrator must first aggregate all origin-destination populations currently located in Patch~$q$ to form $\widetilde{x}^{(q)}$ according to~\cref{eq:explicit-populations} before the local intensities $D^{(q)}(\widetilde{x}^{(q)},t)$ can be evaluated. With an explicit RK scheme, the stage-aligned construction of Zunker et al.~\cite{ZUNKER2027_stage_aligned} only advances the $N_PN_C$ patch aggregates of~\cref{eq:explicit-model-zunker}. The patch-specific stage intensities obtained this way are then reused to reconstruct the $N_EN_C$ traveler states in~\cref{eq:explicit-event-update} inside the RK scheme. The traveler states and their updates are not removed from the model but kept out of the global RK state. Both approaches compute explicit $S$-stage RK steps with coefficients $(a_{s\ell},b_s,c_s)$ from $t_n$ to $t_{n+1}=t_n+h, h>0$, where the open interval $(t_n,t_{n+1})$ does not contain a mobility event. Event times are enforced as step boundaries. 

The stage-aligned scheme starts from aggregate and traveler states satisfying~\cref{eq:explicit-populations} and first advances the patch aggregates with the same RK method for $q\in\mathcal{P}$:
\begin{align}
\begin{aligned}
    \widetilde{x}^{(q),(s)}
    &=
    \widetilde{x}^{(q)}(t_n) +h\sum_{\ell=1}^{s-1}
    a_{s\ell}\widetilde{k}^{(q),(\ell)},\\
    D^{(q),(s)}
    &=
    D^{(q)}
    \left(
        \widetilde{x}^{(q),(s)},
        t_n+c_sh
    \right),\\
    \widetilde{k}^{(q),(s)}
    &=
    BD^{(q),(s)} \widetilde{x}^{(q),(s)},
    \qquad\qquad s=1,\ldots,S,\\
    \widetilde{x}^{(q)}(t_{n+1})
    &=
    \widetilde{x}^{(q)}(t_n)+ h\sum_{s=1}^{S} b_s\widetilde{k}^{(q),(s)}.
    \end{aligned}
    \label{eq:aggregate-rk}
\end{align}
Then, the transition intensity matrices $D^{(q),(s)}$ evaluated in these aggregate stages are reused to advance each traveler state $(p,q)\in\mathcal{E}$:
\begin{align}
\begin{aligned}
    x^{(p;q),(s)}
    &=
    x^{(p;q)}(t_n)+h\sum_{\ell=1}^{s-1}a_{s\ell}k^{(p;q),(\ell)},\\
    k^{(p;q),(s)}
    &=
    BD^{(q),(s)}x^{(p;q),(s)},\\
    x^{(p;q)}(t_{n+1})
    &=
    x^{(p;q)}(t_n)+h\sum_{s=1}^{S}b_sk^{(p;q),(s)}.
    \end{aligned}
    \label{eq:stage-aligned-rk}
\end{align}
Only the traveler states $x^{(p;q)}$, $(p,q)\in\mathcal{E}$, i.e., the off-diagonal entries of $\mathcal{X}$ in~\cref{eq:explicit-od-matrix}, are stored and updated.
The diagonal state, representing residents currently in their home patch, can be reconstructed by
\begin{align}
    x^{(q;q)}(t) = \widetilde{x}^{(q)}(t) - \sum_{p:(p,q)\in\mathcal{E}}  x^{(p;q)}(t),
    \qquad q\in\mathcal{P}.
    \label{eq:explicit-home-reconstruction}
\end{align}

If the same RK tableau and time steps are used, direct global integration and the stage-aligned realization in~\cref{eq:aggregate-rk,eq:stage-aligned-rk} are algebraically equivalent. Their numerical results can differ only through floating-point rounding errors, and both retain the same convergence order; cf.~\cite[Theorem~4.4]{ZUNKER2027_stage_aligned}.

\subsubsection*{C) The age-stratified SEIR metapopulation model with explicit mobility}

Following the implicit mobility example, consider local SEIR dynamics as in~\cref{sec:local-model}B and alternating departure-return mobility events with~\cref{eq:explicit-event-fractions}. Each origin-destination state is given by
\begin{align*}
    x^{(p;q)}(t)=\bigl(S_1^{(p;q)}(t),E_1^{(p;q)}(t),I_1^{(p;q)}(t),R_1^{(p;q)}(t),\ldots,S_{N_G}^{(p;q)}(t),E_{N_G}^{(p;q)}(t),I_{N_G}^{(p;q)}(t),R_{N_G}^{(p;q)}(t)\bigr)^T.
\end{align*}
The aggregate state $\widetilde{x}^{(q)}$ in~\cref{eq:explicit-populations} includes, e.g., $\widetilde{I}_j^{(q)}=\sum_{p:(p,q)\in\mathcal{E}_+}I_j^{(p;q)}$ and the population sizes $\widetilde{N}_j^{(q)}$ for a given Patch~$q$.
As the dynamics are population-conserving by construction, the total population sizes $\widetilde{N}_j^{(q)}$ in a Patch~$q$ are constant between mobility events. The interaction-dependent force of infection is calculated by
\begin{align*}
    \lambda_i^{(q)}(t)=\rho_i\sum_{j=1}^{N_G}\phi_{ij}\frac{\widetilde{I}_j^{(q)}(t)}{\widetilde{N}_j^{(q)}}.
\end{align*}
In contrast to the implicit model formulation, the mobility is not explicitly considered in the transition intensities, but only at the mobility event updates. However, compare the similar structure here and in~\cref{eq:Uqj}. The entries in different columns of $\mathcal{X}$ evolve independently of each other. Due to homogeneous mixing, only the forces of infection $\lambda_i^{(q),(s)}$, i.e., $N_G$ values per patch and stage, are required to be passed to the travelers in~\cref{eq:stage-aligned-rk}. As before, the parameters $\rho_i$, $\phi_{ij}$, $T_{E_i}^{-1}$, and $T_{I_i}^{-1}$ are identical in all patches.

\FloatBarrier

\section{Complexity analysis}
\label{sec:complexity}

A complexity analysis is essential for evaluating beforehand whether an algorithm's runtime and memory requirements scale appropriately with data volume to meet the time and resource constraints of natural disaster response. In this section, we quantify the arithmetic work $\mathcal{W}$ and the memory requirements $\mathcal{M}$ of implicit and explicit mobility representation. For the model using explicit Lagrangian mobility, we consider the stage-aligned RK scheme~\cite{ZUNKER2027_stage_aligned}. Updating all traveler and home states at a mobility event (cf.~\cref{eq:explicit-event-delta,eq:explicit-event-update,eq:explicit-event-conservation}) costs $\mathcal{O}((N_P+N_E)N_C)$ operations. In particular with multiple RK steps and stages between mobility events, this cost is negligible. Therefore, we consider a single fixed-step explicit RK update. In~\cref{tab:mobility_taxonomy}, we summarize the considered models and realizations with the corresponding state organization. For completeness, we provide the detailed results for the direct global integration in~\ref{sec:direct-global-complexity}.

\begin{table}[htbp]
\centering
\small
\setlength{\tabcolsep}{5pt}
\renewcommand{\arraystretch}{1.15}
\begin{tabularx}{\textwidth}{@{}
    >{\raggedright\arraybackslash}p{0.20\textwidth}
    >{\raggedright\arraybackslash}p{0.23\textwidth}
    >{\centering\arraybackslash}p{0.22\textwidth}
    >{\raggedright\arraybackslash}X
@{}}
\toprule
\textbf{Mobility representation} &
\textbf{Computational realization} &
\textbf{\shortstack{Global integrator state size}} &
\textbf{Considered populations} \\
\midrule
Implicit &
Fraction-weighted intensity coupling &
$N_PN_C$ &
$N_PN_C$ resident values. \\
Explicit Lagrangian &
Direct global integration &
$(N_P+N_E)N_C$ &
$(N_P+N_E)N_C$ origin-destination values. \\
Explicit Lagrangian &
Stage-aligned solution scheme &
$N_PN_C$ &
$N_PN_C$ patch-aggregated and $N_EN_C$~traveler values. \\
\bottomrule
\end{tabularx}
\caption{\textbf{Mobility representations and their computational realizations, global integrator state size and considered populations.} The last two rows describe the same explicit Lagrangian model with different computational realizations.}
\label{tab:mobility_taxonomy}
\end{table}

We quantify arithmetic work as the number of floating point operations per RK step, while memory usage is measured by the number of stored scalar values. We follow a hybrid approach where we use asymptotic $\mathcal{O}$-notation and neglect small constants but consider all relevant work and memory contributions with their complexities. Although we are fully aware that different operations (e.g., addition or division) do not correspond to the same number of cycles on an actual machine, we consider a simplified setting where all operations contribute equally. The resulting number of operations still provides a fair idea of the algorithms' cost. We furthermore do not distinguish between integers and values in floating point arithmetic. The practical memory usage in bytes can then fairly be approximated with the utilized precisions.

Output histories, local-model parameters, and the prescribed mobility schedule are excluded from the estimates below. For the mobility matrix $M$, we have storage requirements of $\mathcal{O}(N_P+N_E)$ elements. If the graph is dense, the matrix is stored in dense format. If the graph is sparse, a sparse matrix format realizes the asymptotic storage.

\begin{remark}
In this work, we restricted the mobility coefficients to be compartment-independent (cf. Remark~\ref{rem:mobcoeff} and~\cref{eq:explicit-event-fractions}). If we allow the mobility coefficients to be compartment-specific, only the particular multiplications change, not the number of operations. This means that, while the parallel performance might change, the arithmetic complexity does not. However, the particular memory usage increases from $\mathcal{O}(N_E)$ to $\mathcal{O}(N_EN_C)$.
\end{remark}

To simplify the complexity analysis, let us consider an $S$-stage explicit RK scheme that keeps all $S$ stage derivatives $k^{(s)}$, $s=1,\ldots,S$, in memory. While the optimized RK routines in~\cref{sec:parallel-implementation} reduce the memory footprint by reusing stage storage and accumulating the final update in place, this optimization only reduces the memory constants but does not alter the asymptotic memory scaling in the problem dimensions. In contrast to the memory optimization, the stage-aligned implementation yields a larger constant by additionally buffering the transition intensities for each RK stage over several consecutive time steps to advance the traveler states; see~\cref{sec:parallel-stage-aligned}.

Let $\mathcal{C}_D(N_C,N_T)$ denote the number of floating point operations required to evaluate all $N_T$ per-capita transition intensities $d_r^{(p)}$, i.e., the nonzero entries of $D^{(p)}$, for a single patch $p\in\mathcal{P}$. Given these intensities, the sparsity of $B$ and $D$ in~\cref{eq:local-flow,eq:local-rate-matrix} gives $\mathcal{W}(BDx)=\mathcal{O}(N_C+N_T)$ operations. To form the stage state at RK stage $s$, the $s-1$ stage derivatives from the preceding stages must be combined. Over all $S$ stages this amounts to $S(S-1)/2=\mathcal{O}(S^2)$ vector operations, each of the size of the integrated state, e.g., $N_PN_C$ for implicit mobility.

\subsection{Implicit mobility}
\subsubsection{Arithmetic work}
At each RK stage, (i) computing the populations present in the patches, $\widetilde{X}=M^T X$, using~\cref{eq:implicit-patch-aggregation} costs $\mathcal{O}((N_P+N_E)N_C)$ operations, (ii) evaluating the $N_P$ home-phase and $N_P$ mobility-phase transition-intensity terms in~\cref{eq:implicit-patch-intensities} costs $\mathcal{O}(N_P\mathcal{C}_D(N_C,N_T))$, (iii) mapping the patch-specific intensities back to the residents' home patches as described in~\cref{eq:implicit-intensity-aggregation} costs $\mathcal{O}((N_P+N_E)N_T)$, and (iv) combining the three intensity terms in~\cref{eq:implicit-effective-intensity} and evaluating the right-hand side of~\cref{eq:implicit-model} costs $\mathcal{O}(N_P(N_C+N_T))$, since for each patch, the $N_T$ transitions contribute to a derivative vector with $N_C$ entries. In addition, (v) the RK stage combinations cost $\mathcal{O}(S^2N_PN_C)$ per step. Consequently, the arithmetic work of one $S$-stage RK step is
\begin{align}
    \mathcal{W}_{\mathrm{impl}}
    =\mathcal{O}\left(
        \begin{aligned}
        &\underbrace{S(N_P+N_E)N_C}_{\text{(i)}}
        +\underbrace{SN_P\mathcal{C}_D(N_C,N_T)}_{\text{(ii)}}
        +\underbrace{S(N_P+N_E)N_T}_{\text{(iii)}}
        +\underbrace{SN_P(N_C+N_T)}_{\text{(iv)}}
        +\underbrace{S^2N_PN_C}_{\text{(v)}}
        \end{aligned}
    \right).
    \label{eq:implicit-complexity}
\end{align}
The two aggregation terms (i) and (iii) sum to $(N_P+N_E)(N_C+N_T)$. Since $N_P\leq N_P+N_E$ and $N_C\leq N_C+N_T$, this term also bounds (iv) and (v). Therefore,~\cref{eq:implicit-complexity} reduces to $\mathcal{O}(N_P\mathcal{C}_D(N_C,N_T)+(N_P+N_E)(N_C+N_T))$. 
When we assume a fixed local model, then $N_C$, $N_T$, and $\mathcal{C}_D$ are constants and the bounds are linear in $N_P+N_E$. However, for modestly sized $N_P$ and $N_E$, also the number of age groups, increasing $N_C$ and $N_T$, can represent a relevant contribution to the effective workload.

\subsubsection{Memory usage}
 (i) The resident states in~\cref{eq:implicit-model} require $N_PN_C$ values. (ii) The patch-state temporary array $\widetilde{X}$ in~\cref{eq:implicit-patch-aggregation} requires another $N_PN_C$ values. (iii) Using the sparse transition structure in~\cref{eq:local-rate-matrix}, the intensities represented by~\cref{eq:implicit-effective-intensity,eq:implicit-intensity-aggregation} require $\mathcal{O}(N_PN_T)$ values. (iv) Under the assumed full-storage RK realization, the integrator retains $S$ stage derivatives $k^{(s)}$ of length $N_PN_C$. Finally, (v) sparse or dense storage of $M$ from~\cref{eq:implicit-mobility-fractions} requires $\mathcal{O}(N_P+N_E)$ fractions and matrix indices. Hence,
\begin{align}
    \mathcal{M}_{\mathrm{impl}}
    =\mathcal{O}\left(
        \begin{aligned}
         \underbrace{N_PN_C}_{\text{(i)}}
        +\underbrace{N_PN_C}_{\text{(ii)}}
        +\underbrace{N_PN_T}_{\text{(iii)}}
        +\underbrace{SN_PN_C}_{\text{(iv)}}
        +\underbrace{N_P+N_E}_{\text{(v)}}
        \end{aligned}
    \right)
    \label{eq:implicit-memory}
\end{align}
For fixed $S$, (iv) is of the same order as (i) and (ii), such that $\mathcal{M}_{\mathrm{impl}}=\mathcal{O}(N_E+N_P(N_C+N_T))$.

\subsection{Explicit Lagrangian mobility with stage-aligned solution scheme}
\label{sec:stage-aligned-complexity}

\subsubsection{Arithmetic work}
At each RK stage, (i) evaluating the transition intensity matrices $D^{(q),(s)}$ in~\cref{eq:aggregate-rk} for the $N_P$ patch aggregates costs $\mathcal{O}(N_P\mathcal{C}_D(N_C,N_T))$. The precomputed intensities are then reused as (ii) evaluating the aggregate right-hand sides $\widetilde{k}^{(q),(s)}$ in~\cref{eq:aggregate-rk} costs $\mathcal{O}(N_P(N_C+N_T))$, and (iii) evaluating the $N_E$ traveler right-hand sides $k^{(p;q),(s)}$ in~\cref{eq:stage-aligned-rk} costs $\mathcal{O}(N_E(N_C+N_T))$ without reevaluating the transition intensities.
Finally, (iv) forming all stage states and RK combinations in~\cref{eq:aggregate-rk,eq:stage-aligned-rk} costs $\mathcal{O}(S^2(N_P+N_E)N_C)$ over the complete step. Consequently, the arithmetic work of one $S$-stage step is
\begin{align}
    \mathcal{W}_{\mathrm{ex,stage}}
    &=\mathcal{O}\left(
        \begin{aligned}
        \underbrace{SN_P\mathcal{C}_D(N_C,N_T)}_{\text{(i)}}
        +\underbrace{SN_P(N_C+N_T)}_{\text{(ii)}}
        +\underbrace{SN_E(N_C+N_T)}_{\text{(iii)}}
        +\underbrace{S^2(N_P+N_E)N_C}_{\text{(iv)}}
        \end{aligned}
    \right).
    \label{eq:stage-aligned-complexity}
\end{align}
The right-hand side terms (ii) and (iii) combine to $(N_P+N_E)(N_C+N_T)$, which also bounds (iv) up to constant factors. Therefore,~\cref{eq:stage-aligned-complexity} simplifies to $\mathcal{O}\left(N_P\mathcal{C}_D(N_C,N_T)+(N_P+N_E)(N_C+N_T)\right)$.

\subsubsection{Memory usage}

(i) The patch aggregates in~\cref{eq:aggregate-rk} require $N_PN_C$ values, and (ii) the traveler states in~\cref{eq:stage-aligned-rk} require $N_EN_C$ values. The scheme stores (iii) the $S$ aggregate stage derivatives $\widetilde{k}^{(q),(s)}$ requiring $SN_PN_C$ values, and (iv) the patch-specific stage intensities, reused by the travelers, requiring $SN_PN_T$ values. After these intensities have been computed, the travelers can be processed independently, such that (v) their stage derivatives $k^{(p;q),(s)}$ only require a workspace of $SN_C$ values that is reused for every edge. Note that in the later parallel realization, each thread requires its own workspace, which adds $TSN_C$ values for $T$ threads. The mobility coefficients (and potential sparse indices) (vi) require $\mathcal{O}(N_P+N_E)$ values. The home states from~\cref{eq:explicit-home-reconstruction} are reconstructed as needed and need not be stored permanently. Hence,
\begin{align}
    \mathcal{M}_{\mathrm{ex,stage}}
    &=\mathcal{O}\left(
        \begin{aligned}
        \underbrace{N_PN_C}_{\text{(i)}}
        +\underbrace{N_EN_C}_{\text{(ii)}}
        +\underbrace{SN_PN_C}_{\text{(iii)}}
        +\underbrace{SN_PN_T}_{\text{(iv)}}
        +\underbrace{SN_C}_{\text{(v)}}
        +\underbrace{N_P+N_E}_{\text{(vi)}}
        \end{aligned}
    \right)
    \label{eq:stage-aligned-memory}
\end{align}
This simplifies to $\mathcal{M}_{\mathrm{ex,stage}}=\mathcal{O}((N_P+N_E)N_C+SN_P(N_C+N_T))$. Storing every traveler stage derivative explicitly would add $\mathcal{O}(SN_E N_C)$ values on top. By processing the travelers one at a time and reusing the workspace as in (v), also the parallel stage-aligned scheme reduces this bound to $\mathcal{O}(\min(T,N_E)SN_C)$.

\subsection{Operation counts for the SEIR metapopulation model}
\label{sec:operation-counts}

\begin{figure}[H]
    \centering
    \includegraphics[width=\textwidth]{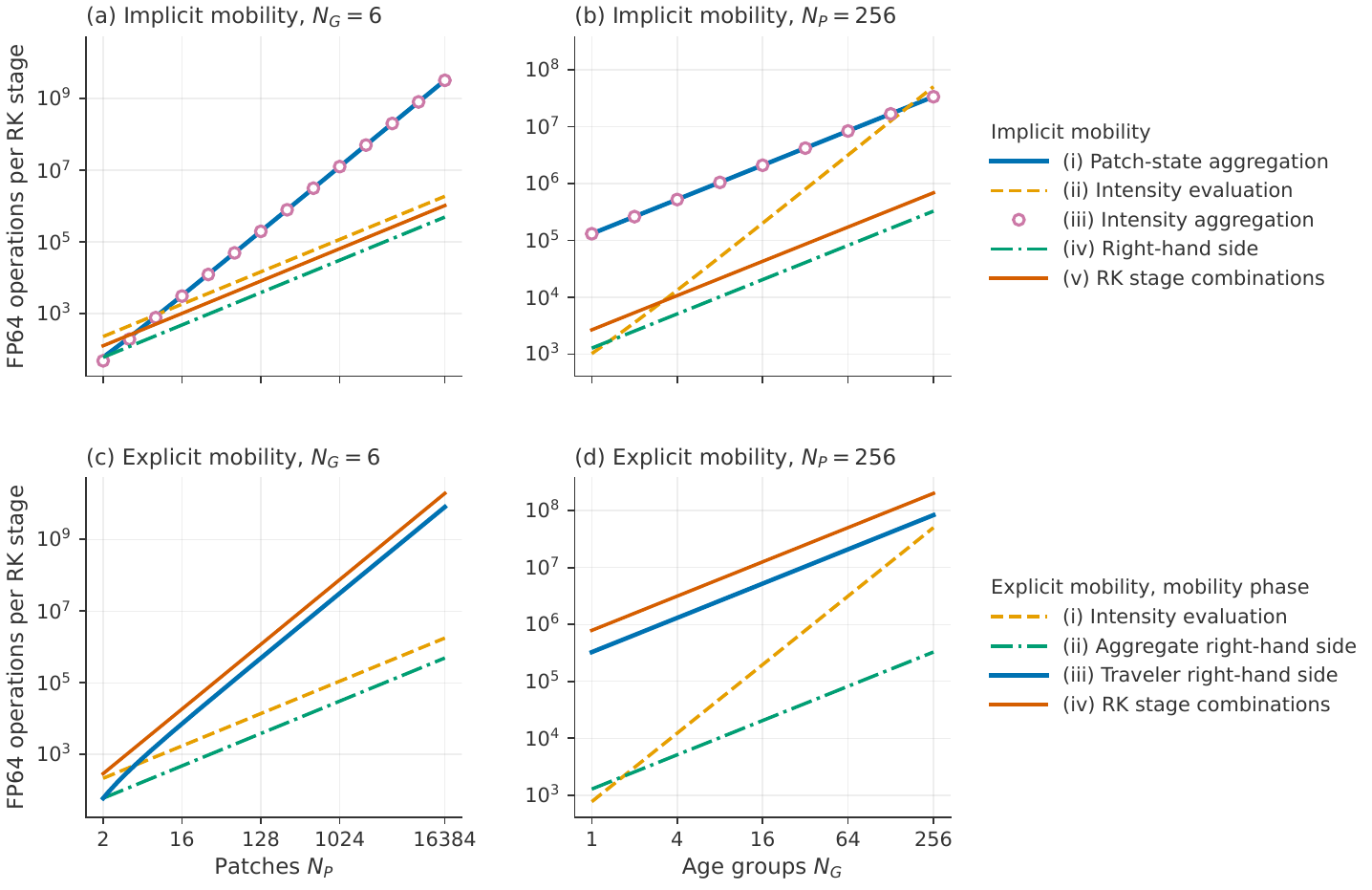}
    \caption{\textbf{Number of FP64 operations in each complexity term.} Number of operations per RK stage for each term in~\cref{eq:implicit-complexity} (implicit mobility (top)) and in~\cref{eq:stage-aligned-complexity} (mobility phase explicit mobility (bottom)), while varying the number of patches for $N_G=6$ (left) and the number of age groups for $N_P=256$ (right).}
    \label{fig:operation-counts}
\end{figure}

In order to visualize the complexity terms in~\cref{eq:implicit-complexity,eq:stage-aligned-complexity}, we quantify the number of floating point operations at each RK stage for each term in our implementation of the age-structured SEIR model, assuming a classical explicit fourth-order Runge-Kutta method (RK4) with $S=4$ stages, see~\ref{app:rk-tableaus}, and a fully connected network, i.e., $N_E=N_P(N_P-1)$. The coefficients for each complexity term are derived in~\ref{app:operation-counts}. As shown in~\cref{fig:operation-counts}, the terms containing the quadratic factor $N_P^2$ exceed all other terms beyond ten patches. On the other hand, only the intensity evaluation grows quadratically with $N_G$. Compared with any aggregation, this requires approximately $(3N_G+1)/(2N_P)$ times as many operations and therefore outperforms the aggregations only for $N_G>(2N_P-1)/3$, e.g., when there are more than 170 age groups for $N_P=256$. Nevertheless note that this evaluation is only preliminary as the corresponding roofline model might identify selected routines as memory-bound. Then, subroutines with less arithmetic work can disproportionately contribute to the algorithm's runtime.

\paragraph{Summary} For fixed $S$, implicit mobility and the stage-aligned scheme share the same leading arithmetic bound $\mathcal{O}(N_P\mathcal{C}_D(N_C,N_T)+(N_P+N_E)(N_C+N_T))$. However, the stage-aligned scheme additionally stores $N_EN_C$ traveler values, whereas implicit mobility stores only $\mathcal{O}(N_E)$ mobility fractions.

\section{Parallel implementation}
\label{sec:parallel-implementation}

To compare both parallel implementations under the same local dynamics, we use the age-structured SEIR metapopulation model of~\cref{sec:metapopulation-models} with an either implicit or explicit mobility representation, implemented in C++ within the MEmilio software framework~\cite{bicker2026memilio}. All computations use double precision arithmetic (FP64) and all implementations assume a fully connected network with a dense mobility matrix $M$.

Parallelization for the CPUs is done with shared memory on the node-level using OpenMP~\cite{dagum1998openmp} and using CUDA~\cite{nickolls2008cuda} on the GPU. In all implementations, we use a classical explicit RK4 scheme whose Butcher tableau is given in~\ref{app:rk-tableaus}. Instead of storing all four stage derivatives, we store only one intermediate stage state and one accumulator for the weighted RK4 update. We integrate only the $S_i$, $E_i$, and $I_i$ compartments, as the recovered compartment $R_i$ is reconstructed from the conserved population in age group $i$. The resident states of the implicit model and the patch aggregates of the stage-aligned scheme are stored as flat arrays grouped by compartment, i.e., the value of compartment $c$ in Patch~$q$ is located at index $(c-1)N_P+q$. This layout allows contiguous accesses for CPU SIMD operations and coalesced accesses on the GPU, i.e., neighboring CUDA threads read contiguous addresses that the GPU combines into few memory accesses. Fixing the number of age groups $N_G$ at compile time via template specialization, we enabled unrolling of the age group-specific loops and confirmed the unrolling via the compiler explorer~\cite{godbolt}.

\subsection{Implicit mobility}
\label{sec:parallel-implicit}

We implement the SEIR model of~\cref{sec:implicit-mobility}C with $\theta=1/2$. At RK stage $s$, we obtain $I_j^{(p),(s)}$, denoting the infectious individuals in Patch~$p$ and age group~$j$, from $X_I^{(s)}\in\mathbb{R}^{N_P\times N_G}$. The local dynamics in~\cref{eq:local-seir} conserve, both, each resident age-group population $N_j^{(q)}$ and the population present in each patch, $\widetilde{N}^{(q)}_{j}=\sum_{p=1}^{N_P}m_{pq}N_j^{(p)}$. We can therefore compute once and store $1/N^{(q)}_{j}$ and $1/\widetilde{N}^{(q)}_{j}$. At each RK stage, we compute the stage-specific matrices $U^{(s)}$ from $M^TX_I^{(s)}$ and $V^{(s)}=MU^{(s)}$ according to~\cref{eq:Uqj,eq:implicit-seir-V} and evaluate the force of infection~\cref{eq:implicit-seir-force}. Each row of $V^{(s)} =M U^{(s)}$ depends on all entries of $U^{(s)}$. The implementations below therefore require finishing computing $U^{(s)}$ before $V^{(s)}$ can be computed.

\subsubsection{CPU implementation}

The OpenMP implementation stores both $M$ and $M^T$ with the entries of each row placed consecutively in memory. While this requires $2(N_P+N_E)$ matrix values, the corresponding memory layout allows matrix-vector operations with $M$ and $M^T$ to use the entire cache line of loaded mobility coefficients. A single parallel region covers all $K$ RK4 steps of the integration (see Algorithm~\ref{alg:implicit-openmp}). As the workload is expected to be equal across all patches, loops are scheduled statically over the patches. 

At each stage, the first loop distributes the patches $q$ and the rows of $M^T$ in equally sized blocks over the threads. For every Patch~$q$, the corresponding thread computes the $N_G$ values of $U_{q:}^{(s)}$ and stores them. The automatic barrier at the end of the loop ensures that all of $U^{(s)}$ is available before the next loop starts. The second loop distributes the patches $p$ and the rows of $M$ in the same way. For each Patch~$p$, the corresponding thread computes the $N_G$ values of $V_{p:}^{(s)}$ in local storage and immediately evaluates~\cref{eq:implicit-seir-force}, the SEIR flows, and the RK4 update. 
Since each row $V_{p:}^{(s)}$ is only used by the thread that it is computed by, $V^{(s)}$ is never stored completely. Each stage therefore uses two parallel loops and synchronization is only handled upon finishing the parallel loops.

\begin{algorithm}[H]
\footnotesize
\DontPrintSemicolon
\KwIn{Initial values $S_i^{(p)}(t_0)$, $E_i^{(p)}(t_0)$, $I_i^{(p)}(t_0)$ of all patches and age groups, mobility matrices $M$ and $M^T$, each stored row by row, stored values $1/N_j^{(p)}$ and $1/\widetilde{N}_{j}^{(q)}$, model parameters, step size $h$, and total number of steps $K$}
\textbf{omp parallel}\;
\Indp
\For{$n\gets1$ \KwTo $K$}{
    \For{$s\gets1$ \KwTo $4$}{
        \OMPFOR{$q=1,\ldots,N_P$ \textnormal{with} \texttt{schedule(static)}}{
            Compute $U_{qj}^{(s)}\gets\sum_{r=1}^{N_P}m_{rq}I_{j}^{(r),(s)}/\widetilde{N}^{(q)}_{j}$ for all age groups $j$\;
        }
        \OMPFOR{$p=1,\ldots,N_P$ \textnormal{with} \texttt{schedule(static)}}{
            Compute $V_{p:}^{(s)}\gets\sum_{q=1}^{N_P}m_{pq}U_{q:}^{(s)}$ in thread-local storage\;
            Evaluate $\lambda_i^{(p),(s)}$ for all $i$, the SEIR derivatives, and the RK4 update\;
        }
    }
}
\Indm
\caption{OpenMP implementation of $K$ RK4 steps for implicit mobility.}
\label{alg:implicit-openmp}
\end{algorithm}

\subsubsection{GPU implementation}

All state, parameter, and working arrays remain on the GPU throughout integration and no transfer takes place between the host and the device. As shown in Algorithm~\ref{alg:implicit-cuda}, each RK stage uses the products of the $N_P\times N_P$ matrices $M^T$ and $M$ with one matrix of size $N_P\times N_G$ via cuBLAS routines and two individual CUDA kernels. The first product computes $M^T X_I^{(s)}$, and the first kernel divides its $N_PN_G$ entries element-wise by the corresponding entry of $\widetilde{N}$. The second product computes $MU^{(s)}$. The final kernel assigns one thread to each patch and evaluates~\cref{eq:implicit-seir-force}, the SEIR derivatives, and the RK4 update for all age groups. Unlike on the CPU, the entries of $M$ are stored only once to keep the memory footprint small. Since cuBLAS assumes column-wise storage, it interprets the row-wise stored $M$ as $M^T$, and only the second product requires the cuBLAS transpose option.

The fixed coefficients $\rho_i\phi_{ij}/2$ and rates $1/T_{E_i}$ and $1/T_{I_i}$ are stored in CUDA constant memory. All operations use the same CUDA stream and therefore run in the required order without CPU synchronization between stages. The $K$ integration steps of one simulated day are recorded once as a CUDA Graph and then launched once per simulated day, reducing kernel-launch overhead.

\begin{algorithm}[H]
\footnotesize
\DontPrintSemicolon
\KwIn{Initial values $S_i^{(p)}(t_0)$, $E_i^{(p)}(t_0)$, $I_i^{(p)}(t_0)$ of all patches and age groups, mobility matrix $M$, stored values $1/N_j^{(p)}$ and $1/\widetilde{N}_j^{(q)}$, model parameters, step size $h$, and number of steps $K$ per simulated day}
Begin capturing the following GPU operations into a CUDA Graph\;
\For{$n\gets1$ \KwTo $K$}{
    \For{$s\gets1$ \KwTo $4$}{
        Compute $\widetilde{X}_I^{(s)}\gets M^T X_I^{(s)}$ with cuBLAS\;
        Launch the normalization kernel to divide each entry of $\widetilde{X}_I^{(s)}$ by the corresponding entry of $\widetilde{N}$ and store $U^{(s)}$\;
        Compute $V^{(s)}\gets MU^{(s)}$ with cuBLAS\;
        Launch the patch kernel to evaluate $\lambda_i^{(p),(s)}$ for all $i$ and $p$, the SEIR derivatives, and the RK4 update\;
    }
}
End the capture and instantiate the CUDA Graph\;
Launch the CUDA Graph once per simulated day and wait for completion after the last day\;
\caption{CUDA implementation of $K$ RK4 steps for implicit mobility.}
\label{alg:implicit-cuda}
\end{algorithm}

\subsection{Stage-aligned solving scheme}
\label{sec:parallel-stage-aligned}

For the explicit Lagrangian setting, we consider distinct home and mobility phases, where during the home phase, all subpopulations are in their home patches. The off-diagonal traveler arrays are inactive during this phase and are overwritten at the next departure. The products $\rho_i\phi_{ij}$ entering the force of infection in~\cref{eq:local-foi} are held constant and stored in an $N_G\times N_G$ infection-coefficient matrix.

Between mobility events, the aggregate dynamics in~\cref{eq:aggregate-rk} are decoupled and we therefore process all integration steps until the next event. Let $K$ be the number of steps up to the next mobility event and $L_{\max}=64$ the maximum number processed in one batch. Each batch contains $L\leq L_{\max}$ steps and never crosses an event. First, the patch aggregates are advanced for all $L$ steps, storing the forces of infection $\lambda_i^{(q),(k,s)}$ for step $k=1,\ldots,L$, stage $s=1,\ldots,4$, and age group $i=1,\ldots,N_G$. The batch-specific storage only contains the forces of infection, i.e., $4N_G$ values per step, bounding the batch by $L_{\max}$ should ensure that the stored values reside in a lower cache level. With $L_{\max}=64$, every mobility phase of our benchmarks, with at most 32 steps, is processed in a single batch, while the history of one patch remains small, i.e., about 12~KB for $N_G=6$. The traveler states then undergo all four stages of~\cref{eq:stage-aligned-rk} for each of these steps, using their stored forces of infection $\lambda_i^{(q),(k,s)}$ from their current Patch~$q$. Thus, every traveler still receives a complete RK4 update at every step. Computations are independent across patches and, once their aggregate histories are available, across travelers and age groups. 

Aggregate values use a flat vector layout with indexing $(c-1)N_P+q$, where $c=1,\ldots,N_C$. Traveler values are stored in a flat layout with indexing $((q-1)N_C+c-1)(N_P-1)+m$, where $q\in\mathcal{P}$ is the destination patch in which the travelers are currently located, and $m=1,\ldots,N_P-1$ denotes the index of the traveler states $x^{(p;q)}$ from the other patches $p\neq q$. Thus, for a fixed destination patch and compartment, the values of all travelers are contiguous. The mapping to the origin $p\ne q$ is not required during the integration phase. Intermediate RK4 states and accumulators remain local to the updating routine or thread. Only the forces of infection are retained for all steps in a batch. Their storage differs between CPU and GPU as described below. Aggregate and traveler states are updated in place.

\subsubsection{CPU implementation}

A single OpenMP parallel region spans the complete simulation, including its mobility events. As the workload is expected to be identical across all patches, we use static \texttt{omp for} scheduling to distribute patches across the threads. As outlined in Algorithm~\ref{alg:stage-aligned-openmp}, each assigned thread advances its patch through all $K$ steps up to the next event. For each batch, it first computes the $L$ aggregate steps and copies their stage-specific forces of infection into a history with $4L_{\max}N_G$ values, i.e., one value for each of the four RK4 stages, $L_{\max}$ steps, and $N_G$ age groups. 
Within this history, the four stage-specific force of infection values for each RK-step are stored contiguously, and all steps of an age group form one contiguous block, i.e., the value for age group $i$, step $k$, and stage $s$ is located at index $((i-1)L_{\max}+k-1)4+s$. The same thread then updates the patch's travelers from this history. No barrier between the aggregate and traveler calculations is needed within a patch, since no other thread accesses its history. The automatic barrier at the end of the patch loop ensures that all patches have reached the next mobility event.

For each age group, the thread processes consecutive segments of at most $B_{\mathrm{CPU}}=256$ traveler indices with $N_C/N_G=4$ state values of FP64 precision such that the resulting segment of 8~KB and the history of 12~KB (for $N_G=6$) fit into an L1 cache of 48~KB. There are $n_{\mathrm{CPU}}=\lceil(N_P-1)/B_{\mathrm{CPU}}\rceil$ segments for each patch and each age group, i.e., $N_Gn_{\mathrm{CPU}}$ segments per patch, with a possibly shorter final segment. All $L$ steps are applied to one segment before moving to the next, allowing the traveler values of the segment to be reused in cache. For each step, the four stored forces of infection and two independent transition rates are loaded before the innermost loop. The innermost loop over the traveler index $m$ is marked with \texttt{omp simd}, instructing the compiler to vectorize it. Since all travelers of a segment have the same forces of infection and rates (homogeneous mixing) and their values are stored contiguously, one SIMD instruction can update several travelers at once.

At departure events, a parallel loop over the destination patches $q$ computes the traveler states $x^{(p;q)}=m_{pq}x^{(p;p)}$ of all origins and the aggregates $\widetilde{x}^{(q)}$ from the states at the time of departure. At return, a parallel loop over the home patches $p$ computes the local individuals $x^{(p;p)}$ using~\cref{eq:explicit-home-reconstruction} and adds the returning travelers $x^{(p;q)}$ from all other connected patches.

\begin{algorithm}[H]
\footnotesize
\DontPrintSemicolon
\KwIn{Aggregate and traveler arrays, step size $h$, fractions $m_{pq}$, $K$ steps per half day, batch capacity $L_{\max}$, and segment length $B_{\mathrm{CPU}}$}
\textbf{omp parallel}\;
\Indp
\ForEach{simulated day}{
    \tcp{Home phase}
    \OMPFOR{$q=1,\ldots,N_P$ \textnormal{with} \texttt{schedule(static)}}{
        Advance $\widetilde{x}^{(q)}$ by $K$ RK4 steps\;
    }
    \tcp{Departure}
    \OMPFOR{$q=1,\ldots,N_P$ \textnormal{with} \texttt{schedule(static)}}{
        Compute $x^{(p;q)}\gets m_{pq}x^{(p;p)}$ for all $p\neq q$ and $\widetilde{x}^{(q)}\gets\sum_{p=1}^{N_P}m_{pq}x^{(p;p)}$\;
    }
    \tcp{Mobility phase}
    \OMPFOR{$q=1,\ldots,N_P$ \textnormal{with} \texttt{schedule(static)}}{
        $\ell\gets0$\;
        \While{$\ell<K$}{
            $L\gets\min(L_{\max},K-\ell)$\;
            \For{$k\gets1$ \KwTo $L$}{
                Advance $\widetilde{x}^{(q)}$ by one RK4 step and store $\lambda_i^{(q),(k,s)}$ in history for all $i,s$\;
            }
            \ForEach{age group $i$}{
                \ForEach{segment $b$ of at most $B_{\mathrm{CPU}}$ travelers}{
                    \For{$k\gets1$ \KwTo $L$}{
                        Load $\lambda_i^{(q),(k,s)}$, $s=1,\ldots,4$, and the two independent transition rates\;
                        \OMPSIMD{traveler index $m$ in segment $b$}{
                            Perform the four-stage update of $(S_i,E_i,I_i)$ in the traveler state $x^{(p;q)}$ with index $m$; reconstruct $R_i$\;
                        }
                    }
                }
            }
            $\ell\gets\ell+L$\;
        }
    }
    \tcp{Return}
    \OMPFOR{$p=1,\ldots,N_P$ \textnormal{with} \texttt{schedule(static)}}{
        Compute $x^{(p;p)}\gets\widetilde{x}^{(p)}-\sum_{q\neq p}x^{(q;p)}$ by~\cref{eq:explicit-home-reconstruction}\;
        Compute $\widetilde{x}^{(p)}\gets x^{(p;p)}+\sum_{q\neq p}x^{(p;q)}$\;
    }
}
\Indm
\caption{Batched OpenMP implementation of stage-aligned explicit mobility (one day with home phase, departure, mobility phase, and return).}
\label{alg:stage-aligned-openmp}
\end{algorithm}

\subsubsection{GPU implementation}

The CUDA implementation uses two sequential kernel launches per batch of $L$ steps, as outlined in Algorithm~\ref{alg:stage-aligned-cuda}. 

The aggregate kernel uses a one-dimensional grid with one thread per Patch~$q$. Threads within a block cooperatively load the $N_G^2$ infection coefficients $\rho_i\phi_{ij}$ and the $2N_G$ rates $1/T_{E_i}$ and $1/T_{I_i}$ into shared memory before synchronizing. Each thread then advances its patch aggregate through all $L$ steps and writes its $4LN_G$ forces of infection. These are stored in a global history with one value per RK stage, step within the batch, age group, and patch, requiring capacity for $4L_{\max}N_GN_P$ values. Within this history, the values of each patch and age group form one contiguous block, in which the four stage values of each RK step are stored contiguously, i.e., the value for Patch~$q$, age group~$i$, step~$k$, and stage~$s$ is located at index $(((q-1)N_G+i-1)L_{\max}+k-1)4+s$. Thus, a traveler block can load its complete history with coalesced accesses. Aggregate states retain a layout with indexing $(c-1)N_P+q$, which supports coalesced accesses. The aggregate block contains at most $B_A=64$ threads.

The traveler kernel uses a three-dimensional grid with dimensions $(n_T,N_P,N_G)$, where $n_T=\lceil(N_P-1)/B_T\rceil$ and $B_T$ is its thread-block size. The first grid dimension indexes the traveler segment $b$, the second the Patch~$q$, and the third the age group $i$. Each block cooperatively loads the $4L$ forces for its patch and age group, together with the two independent transition rates, into shared memory and then synchronizes. Each thread handles one traveler index $m=bB_T+\texttt{threadIdx.x}+1$ in traveler segment $b$ and age group $i$. It loads the four SEIR values once, performs all $L$ RK steps in the batch using the stored forces of infection $\lambda_i^{(q),(k,s)}$, $s=1,\ldots,4$, and stores the final values once. The values remain in thread-local variables between steps, although their placement in registers or local memory is a compiler decision. The accesses of neighboring threads to the traveler values are coalesced, and all threads in a block reuse the same stage-specific force of infection history. The traveler block contains at most $B_T=256$ threads.

Both kernels run in the same CUDA stream, so the traveler kernel starts only after the aggregate kernel is complete. This ordering is required once per batch, not between every pair of RK steps. As shown in Algorithm~\ref{alg:stage-aligned-cuda}, the simulation also performs the home phase, departure, and return according to~\cref{eq:explicit-event-delta,eq:explicit-event-update} on the GPU, and one complete day is recorded as a CUDA Graph that is launched once per simulated day.

\begin{algorithm}[H]
\footnotesize
\DontPrintSemicolon
\KwIn{Aggregate and traveler arrays, step size $h$, fractions $m_{pq}$, $K$ steps per half day, batch capacity $L_{\max}$, block sizes $B_A$ and $B_T$, and a CUDA stream}
$n_A\gets\lceil N_P/B_A\rceil$ and $n_T\gets\lceil(N_P-1)/B_T\rceil$\;
Begin capturing the following GPU operations into a CUDA Graph\;
\tcp{Home phase}
Launch the aggregate kernel to advance all $\widetilde{x}^{(q)}$ by $K$ RK4 steps without storing $\lambda_i^{(q),(k,s)}$\;
\tcp{Departure}
Launch the departure kernel with one block per patch $q$ and compartment to compute $x^{(p;q)}\gets m_{pq}x^{(p;p)}$ for all $p\neq q$ and $\widetilde{x}^{(q)}\gets\sum_{p=1}^{N_P}m_{pq}x^{(p;p)}$\;
\tcp{Mobility phase}
$\ell\gets0$\;
\While{$\ell<K$}{
    $L\gets\min(L_{\max},K-\ell)$\;
    Launch the aggregate kernel with $n_A$ blocks of $B_A$ threads\;
    \ForEach{aggregate-kernel block in parallel}{
        Cooperatively load infection coefficients and rates, then synchronize the block\;
        \ForEach{patch thread $q$}{
            \For{$k\gets1$ \KwTo $L$}{
                Advance $\widetilde{x}^{(q)}$ by one RK4 step and store $\lambda_i^{(q),(k,s)}$ for all $i,s$\;
            }
        }
    }
    Launch the traveler kernel with grid $(n_T,N_P,N_G)$ and $B_T$ threads per block in the same stream\;
    \ForEach{traveler-kernel block $(b,q,i)$ in parallel}{
        Cooperatively load the $4L$ forces and two independent transition rates, then synchronize the block\;
        $m\gets bB_T+\texttt{threadIdx.x} + 1$\;
        \If{$m\leq N_P-1$}{
            Load $(S_i,E_i,I_i,R_i)$ of the traveler state $x^{(p;q)}$ with index $m$\;
            \For{$k\gets1$ \KwTo $L$}{
                Perform the four-stage update using $\lambda_i^{(q),(k,s)}$ and reconstruct $R_i$\;
            }
            Store the updated $(S_i,E_i,I_i,R_i)$\;
        }
    }
    $\ell\gets\ell+L$\;
}
\tcp{Return}
Launch the stayer kernel to compute $x^{(q;q)}$ by~\cref{eq:explicit-home-reconstruction} for all $q$\;
Launch the return kernel with one thread per patch $p$ and compartment to compute $\widetilde{x}^{(p)}\gets x^{(p;p)}+\sum_{q\neq p}x^{(p;q)}$\;
End the capture and instantiate the CUDA Graph\;
Launch the CUDA Graph once per simulated day\;
\caption{Batched CUDA implementation of stage-aligned explicit mobility (one day with home phase, departure, mobility phase, and return).}
\label{alg:stage-aligned-cuda}
\end{algorithm}

\section{Numerical results}
\label{sec:numerical-results}

For runtime and scaling measurements, CPU runs use two 64-core AMD EPYC 9555 processors (3.2--4.4~GHz). We use one OpenMP thread for serial execution and up to 128 OpenMP threads, each pinned to one physical core, for parallel experiments. For experiments with $N_P<128$ patches, we use $N_P$ many threads. CUDA runs use one 80~GB NVIDIA H100 SXM GPU on a node with an Intel Xeon Gold 6530 CPU. For comparison, we also run the same program on one 80~GB NVIDIA A100 SXM4 GPU on a node with an AMD EPYC 7702 CPU. We use Google Benchmark to record the wall-clock time for all integration, synchronization, and mobility operations, excluding setup, resets, an initial CPU-GPU transfer, and output. Only the final resident state is used as simulation output. We use CUDA~12.4~\cite{nickolls2008cuda} and compile all CPU code with GCC~10.4. 

\subsection{Effect of the mobility representation on epidemic dynamics}
\label{sec:model-outcomes}

Implicit and explicit Lagrangian mobility can use the same mobility fractions while defining different systems. To compare them, we consider a simple scenario with one age group and four equally populated patches arranged in a line. The epidemic starts with exposed and infectious individuals in Patch~1, and mobility is restricted to adjacent patches.
Except for the mobility representation and the mobility cycle lengths in the explicit mobility case, we use the same setting for all simulations; see~\cref{tab:mobility-outcome-setup} for more details.

In the implicit formulation, the home and mobility terms contribute with equal weights, $\theta=1/2$. The formulation does not alternate between home and mobility phases and therefore has no cycle length. In contrast, the explicit mobility model defines the cycles through the sequence $(\tau_a)_{a\in\mathbb{N}_0}$. We consider five explicit simulations with constant cycle lengths of $\delta\in\{0.25,0.5,1,2,4\}$~days. In each explicit schedule, residents spend the first half of each cycle in their home patches and the second half in the mobility phase. For $\delta=0.25$, $0.5$, $1$, $2$, and $4$~days, each phase lasts $3$, $6$, $12$, $24$, and $48$~hours, respectively. For the primary case $\delta=1$~day, residents depart at $\tau_{2a+1}=a+0.5$~days and return at $\tau_{2a+2}=a+1$~days for $a\in\mathbb{N}_0$. At departure, each resident population is distributed to the patches according to~$M$. At return, all travelers move back to their home patches. The event fractions from~\cref{eq:explicit-event-fractions} are specified in~\cref{tab:mobility-outcome-setup}. We aggregate the explicit state by resident patch as $x_{\mathrm{expl}}^{(p)}(t)=\sum_{q:(p,q)\in\mathcal{E}_+}x^{(p;q)}(t)$. We use the same implicit trajectory as the reference for all explicit cycle lengths, isolating the effect of the switching timescale. 

\begin{figure}[!bt]
    \centering
    \includegraphics[width=\textwidth]{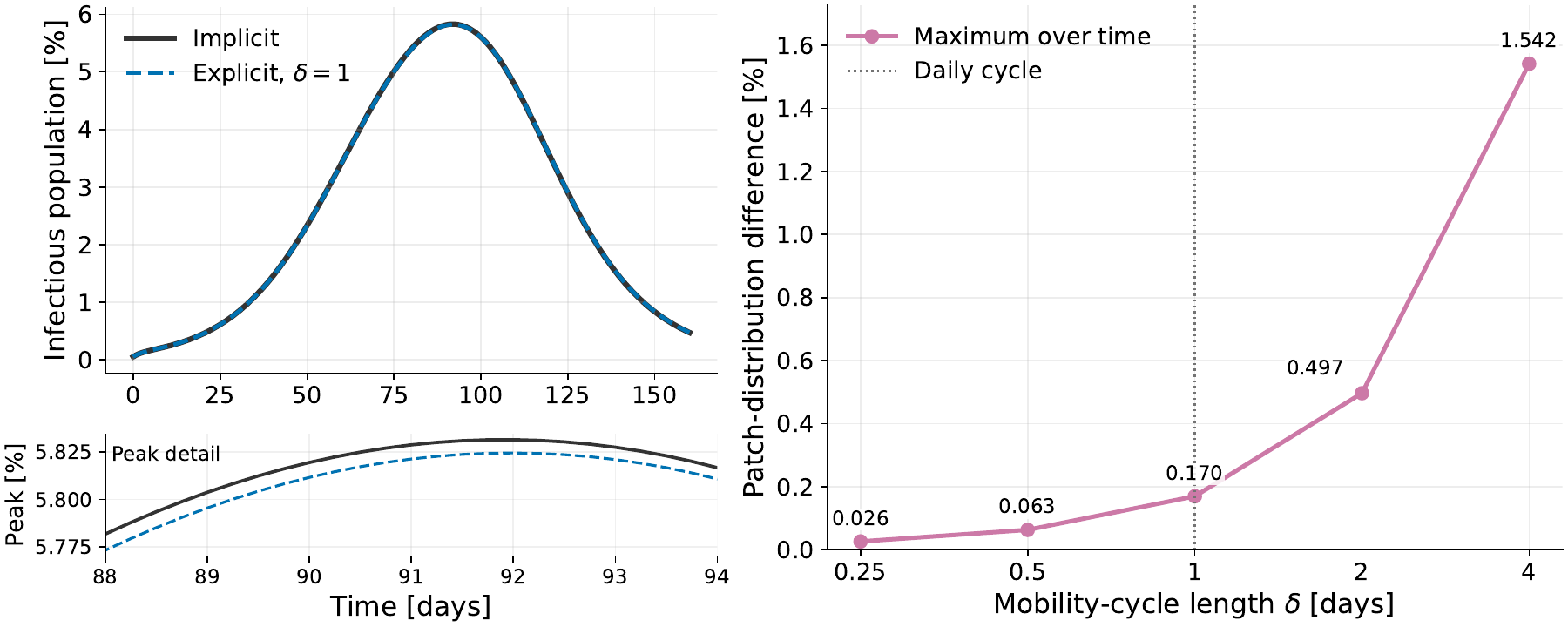}
    \caption{\textbf{Effect of the mobility representation on the simulated epidemic.} The left panel shows the global infectious prevalence for implicit mobility and for the daily explicit cycle; the lower axis enlarges the peak region. The right panel shows the maximum of $\varepsilon_I$ over all output times~$t_n$, i.e., the patch-wise difference normalized by the implicit epidemic peak (see~\cref{eq:mobility-outcome-difference}), as a function of the explicit mobility-cycle length~$\delta$. The dotted line marks the daily cycle used for the primary comparison.}
    \label{fig:mobility-outcomes}
\end{figure}

At the output times $t_n=n/4$~days, i.e., every six hours, we quantify differences between the patch-level infectious populations by
\begin{align}
    \varepsilon_I(t_n,\delta)
    &=\frac{
        \displaystyle\sum_{p\in\mathcal{P}}
        \left|
            I_{\mathrm{expl}}^{(p)}(t_n,\delta)
            -I_{\mathrm{impl}}^{(p)}(t_n)
        \right|
    }{
        \displaystyle\max_m\sum_{p\in\mathcal{P}}
        I_{\mathrm{impl}}^{(p)}(t_m)
    }.
    \label{eq:mobility-outcome-difference}
\end{align}
The numerator is the patch-wise absolute difference, while the denominator is the maximum global infectious population in the implicit simulation. Thus, $\varepsilon_I$ is the patch-wise difference normalized by the implicit epidemic peak. The right panel of~\cref{fig:mobility-outcomes} shows the largest recorded value of~$\varepsilon_I$, expressed as a percentage, for each cycle length~$\delta$.

Before directly comparing the formulations, we validate the results themselves. Without any mobility ($M=I_{N_P}$), the maximum difference between the two formulations is $8.9\times10^{-19}$. The population in each patch is conserved with a relative error below $4.0\times10^{-14}$, and the aggregate explicit state equals the sum of its origin-destination states to within $4.5\times10^{-16}$, i.e., up to rounding errors.

The implicit mobility model reaches a global peak on day~92 with $5.83~\%$ infectious individuals in the population; cf.~\cref{fig:mobility-outcomes}. The daily explicit simulation peaks at the same output time, at $5.82~\%$. For this simulation, $\varepsilon_I$ reaches its maximum of $0.17~\%$ on day~83.75. At day~92, depending on the patch, the explicit simulations with $\delta=0.25$, $0.5$, $1$, $2$, and $4$~days have up to $3.3$, $8.3$, $22.9$, $69.3$, and $222.6$ fewer or up to $0.6$, $1.2$, $2.6$, $5.9$, and $14.3$ more infectious individuals per $100{,}000$ residents than the implicit simulation, respectively. For $\delta=1$, the largest absolute patch-level difference across all patches and output times is $24.8$ per $100{,}000$, which is $0.52~\%$ of the corresponding infectious count in the implicit simulation at that time and patch. Across the tested cycle lengths, the largest recorded value of~$\varepsilon_I$ increases with $\delta$, from $0.026~\%$ for $\delta=0.25$~days to $1.54~\%$ for $\delta=4$~days. In the latter case, the global peak occurs on day~92.5 and reaches $5.76~\%$. Thus, the global trajectories remain close in this scenario, whereas the patch-level discrepancy increases with the duration of the uninterrupted mobility phases.

\begin{figure}[!b]
    \centering
    \includegraphics[width=\textwidth]{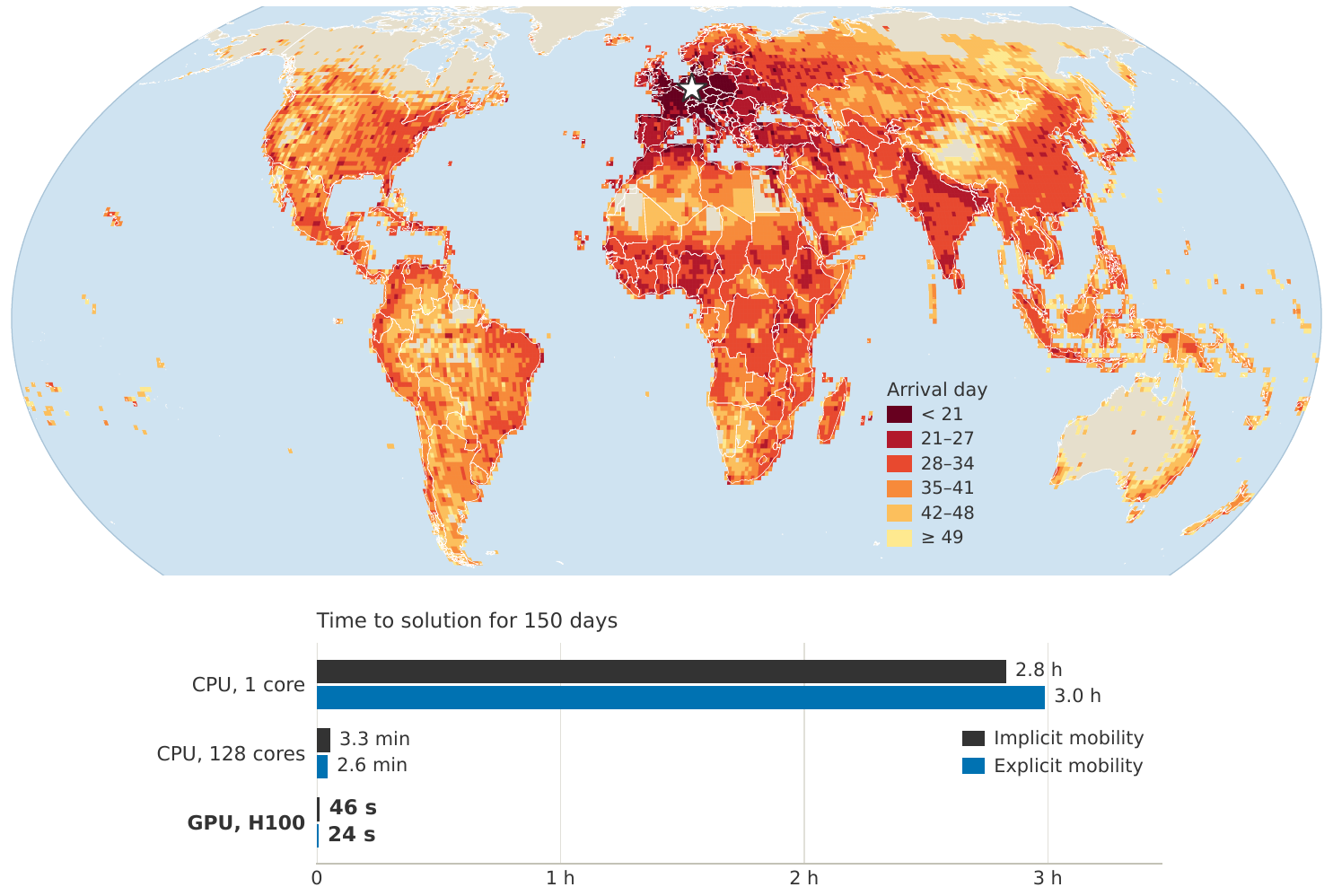}
    \caption{\textbf{Synthesized pandemic application with $\mathbf{N_P=12{,}865}$ patches.} The map view shows the arrival day of the epidemic, i.e., the first day with at least one infectious resident, for explicit mobility. The star marks the initially infected patch of Frankfurt, and land in beige has fewer than 1000 inhabitants per cell. The time to solution for a 150-days simulation with $h=1/64$, on one CPU core, on 128 CPU cores, and on one H100 GPU, is shown below.}
    \label{fig:world-application}
\end{figure}

The close agreement between the trajectories in the left panel and the cycle-length dependence in the right panel of~\cref{fig:mobility-outcomes} can be explained by the duration of the explicit phases.
For short cycles $\delta$, the epidemic state changes little within either phase, so the combined home- and mobility-phase effect approaches the equal weighting used by the implicit model. For longer cycles, the mobility phase starts from a substantially changed state. The nonlinear, patch-dependent transition $S\to E$ makes the phase length relevant, whereas the common linear transitions $E\to I$ and $I\to R$ do not. Several features of the setup limit the differences between the two formulations: the daily cycle is short compared with $T_E=4$~days and $T_I=6$~days, $80-90~\%$ of residents remain in their home patch during the mobility phase, and all patches have equal populations and epidemiological parameters. The result should therefore not be interpreted as a general equivalence of the two mobility representations.

\subsection{Application for a global pandemic use case}
\label{sec:world-application}

Before diving deeper into the performance analysis, we show that our optimized parallel metapopulation models enable the use of large-scale realistic simulations. Therefore, we consider a synthesized global pandemic use case. We use population data from CIESIN's Gridded Population of the World, Version 4~(GPWv4) provided by NASA~\cite{Network-CIESIN-Columbia_University_2018}. We retain all cells of the $1^\circ$ grid for the year 2020 with at least $1000$ inhabitants, which yields $12{,}865$ patches with local populations. 

We generate mobility connections by a simplified gravity model and local disease dynamics are loosely motivated by SARS-CoV-2 dynamics. Each patch sends $10~\%$ of its population to the other patches, distributed proportionally to $N^{(q)}/\xi_{pq}^2$, i.e., a singly-constrained gravity model~\cite{Barbosa_2018}, where $N^{(q)}$ is the population of patch~$q$ and $\xi_{pq}$ denotes the great-circle distance between the cell centers.
The age groups, epidemiological parameters, and mobility events correspond to~\cref{tab:runtime-setup} with $N_G=6$. The epidemic begins with $100$ infectious individuals in the Frankfurt patch. We simulate $150$ days with $h=1/64$. As shown in~\cref{fig:world-application}, the epidemic reaches all patches within $57$ days.

We see that our parallelized metapopulation models bring down the already optimized single-core runtime of approximately three hours to less than one minute. With the simplified setting of implicit mobility representations, we require $1.3$~GB such that a single GPU could simultaneously handle around 60 simulations of varying parameter configurations. For the explicit mobility case, we solve approximately $10{,}000$ time steps with four billion state variables in only 24 seconds. The memory requirements of the explicit simulation are about $33$~GB. With 24~s per simulation, a parameter sweep with 1000 simulations would take less than an hour on a handful of GPUs.

\subsection{Detailed runtime analysis on CPU and GPU}
\label{sec:runtime-comparison}

\begin{figure}[!b]
    \centering
    \includegraphics[width=\textwidth]{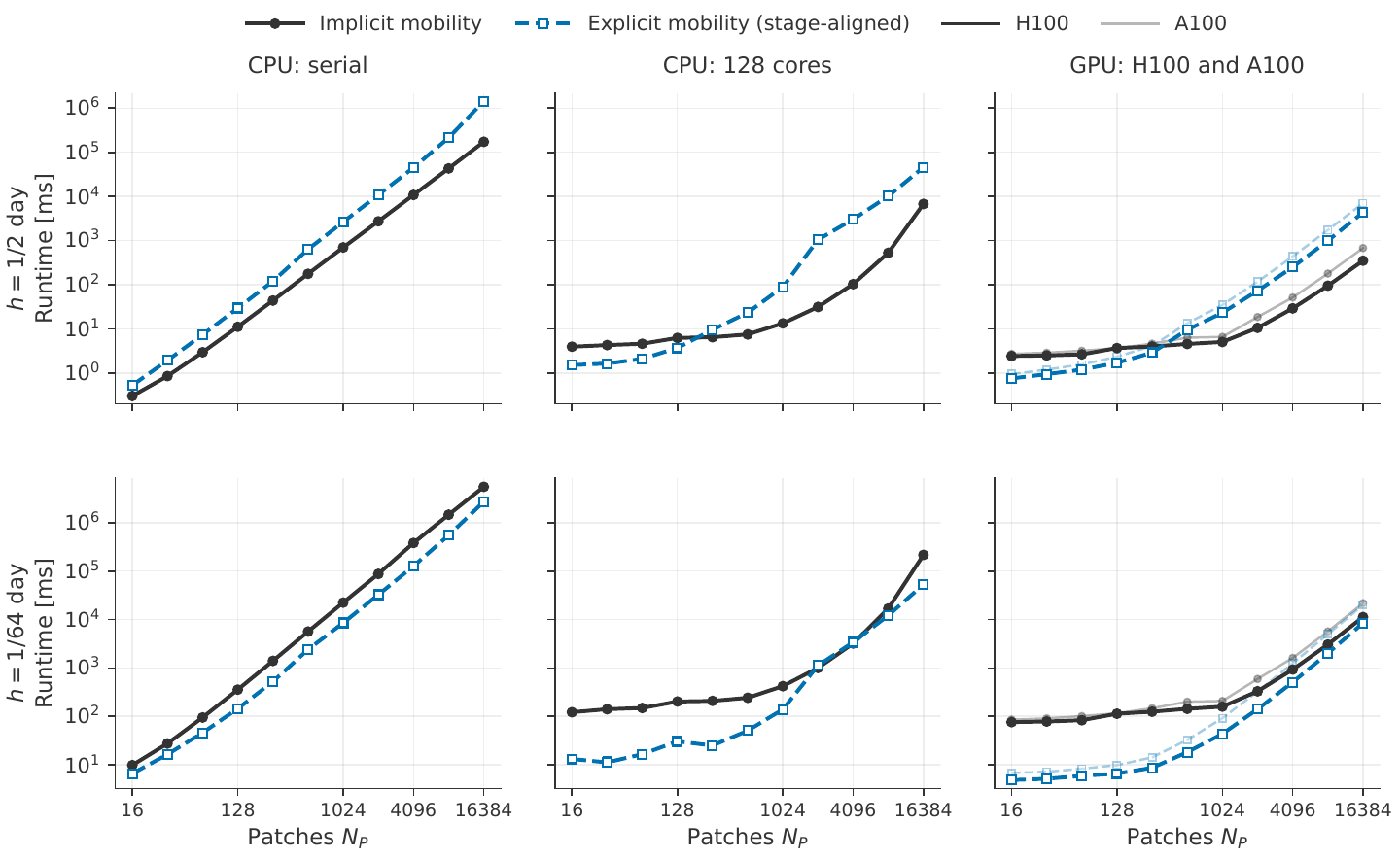}
    \caption{\textbf{Runtime of 32-day simulations with varying $\mathbf{N_P}$ and $\mathbf{N_G=6}$ with serial CPU (left), OpenMP on 128 cores (center), and A100 and H100 GPUs (right).}
    Rows show results with different time steps of $h=1/2$ (top) and $h=1/64$ (bottom). Results are medians of ten repetitions. Timings include explicit mobility events but exclude setup and output.}
    \label{fig:runtime-comparison}
\end{figure}

We now compare both approaches over 32 simulated days with fully connected networks of\linebreak $N_P\in\{16,32,\ldots,8192,16384\}$ with $N_G=6$ and a fixed network of $N_P=1024$ with varying $N_G\in\{1,3,6,8\}$. Populations vary between patches and age groups, and $10-20~\%$ of residents commute; see~\cref{tab:runtime-setup}. 

Figure~\ref{fig:runtime-comparison} shows the results with six age groups ($N_G=6$). Implicit mobility is faster in serial execution if $h=1/2$. Although it evaluates the coupling repeatedly, it does not maintain the origin-specific traveler states that the explicit implementation must advance and redistribute. With only one integration step per half-day, explicit mobility cannot yet exploit data reuse across multiple steps between events. For $h=1/64$, explicit mobility is of the same timing or faster in all tested scenarios.

For small systems, fixed costs per thread synchronization and GPU kernel launch dominate the runtime.
For larger networks and $h=1/2$, implicit mobility is substantially faster than explicit mobility. On the GPU, its advantage increases from a factor of 4.59 at $N_P=1024$ to 12.40 at $N_P=16384$. At the latter size, it reduces runtime from 4.34~s to 0.350~s. On 128 CPU cores, the corresponding largest-case reduction is from 45.64~s to 6.74~s, a factor of 6.77. 
Reducing the step size to $h=1/64$ changes the outcome. Explicit mobility is faster on the GPU but its advantage decreases overall from approximately a factor of 15.7 at $N_P=16$ to 3.63 at $N_P=1024$ and 1.33 at $N_P=16384$. At the largest size, explicit and implicit mobility take 8.44~s and 11.23~s, respectively. On OpenMP, explicit mobility is 4.11 times faster at $N_P=16384$ (52.76~s versus 216.59~s). 

\begin{figure}[!b]
    \centering
    \includegraphics[width=\textwidth]{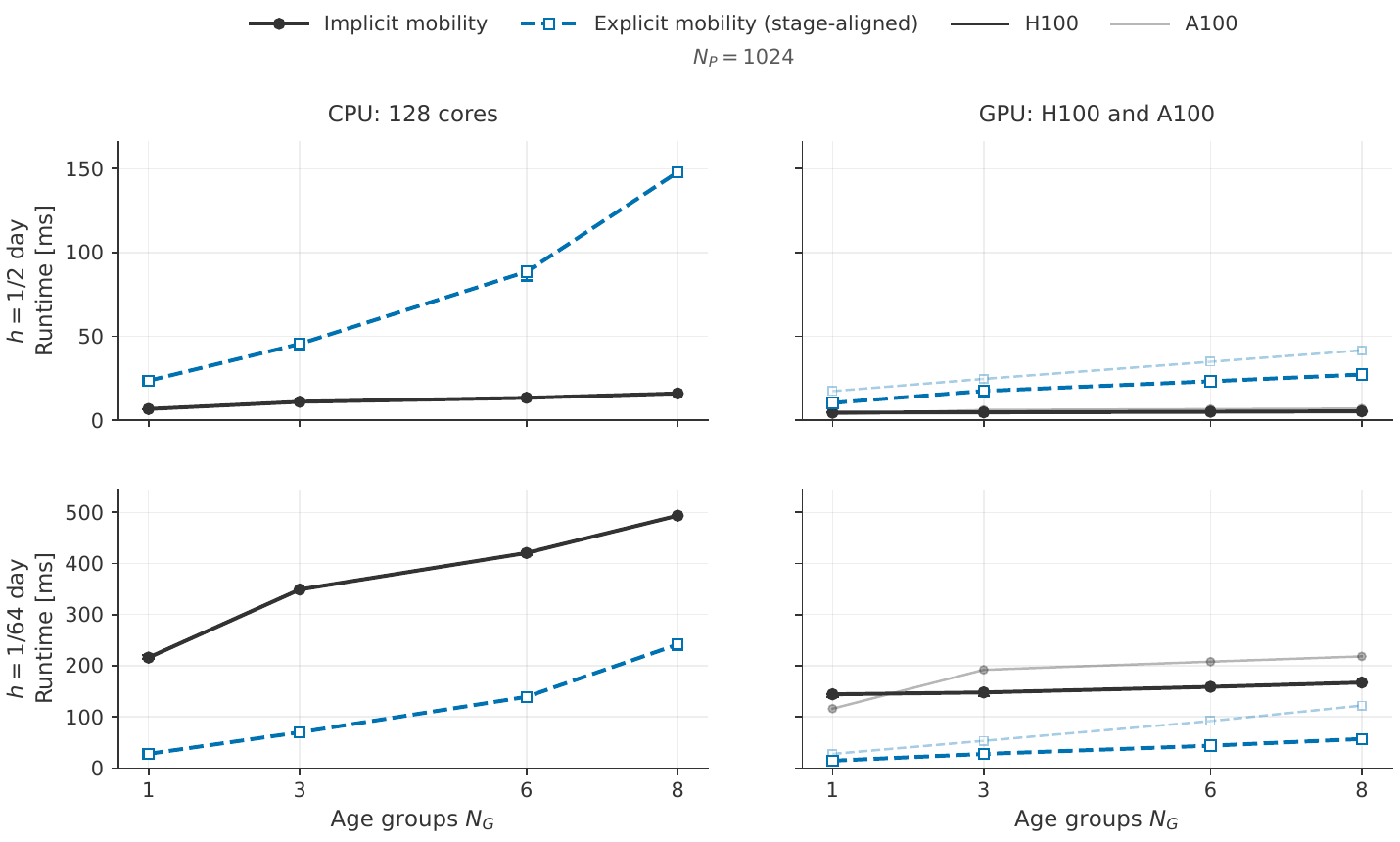}
    \caption{\textbf{Runtime versus the number of age groups at $\mathbf{N_P=1024}$ for 32-day simulations on CPU (left) and GPU (right).} Rows show results with different time steps of $h=1/2$ (top) and $h=1/64$ (bottom). Results are medians of ten repetitions. Timing conditions match those in~\cref{fig:runtime-comparison}.}
    \label{fig:runtime-age-groups}
\end{figure}

The finer step size gives 32 times as many RK4 steps without affecting the 64 explicit mobility events. Implicit mobility repeats its coupled matrix operations at every stage. Explicit mobility advances only resident aggregates during the home phases, when all travelers are at home. Each traveler is therefore advanced for 32 steps over the complete simulation with a step size of $h=1/2$ and 1024 steps over the fine-step simulation. During the mobility phase, every traveler undergoes all four RK4 stages in every step. The batched CPU and GPU implementations in~\cref{sec:parallel-stage-aligned} process all 32 steps of each fine-step half-day together, using the stored forces of infection for each step. This allows reuse of CPU cache and avoids loading and storing the global GPU traveler array at every step. At $N_P=16384$, the finer step size increases implicit runtime by approximately a factor of 32 on both OpenMP and CUDA, compared with factors of 1.16 and 1.94 for explicit mobility. These smaller increases are consistent with data reuse.
Storage requirements might limit the feasible network size: at $N_P=16384$ and $N_G=6$, the explicit traveler array contains $N_EN_C$ FP64 values and occupies 51.5~GB. Doubling $N_P$ would require approximately 206~GB for this array alone, exceeding the 80~GB GPU capacity.

Figure~\ref{fig:runtime-age-groups} considers the age-stratification scaling at $N_P=1024$ for both step sizes. At $h=1/2$, implicit mobility is faster. For $N_G=1$ and $N_G=8$, the factors are 3.51 and 9.26 on 128 CPU cores and 2.26 and 5.12 on the GPU, respectively. At $h=1/64$, stage-aligned solving is faster, but its advantage decreases from a factor of 7.76 to 2.04 on the CPU and from 9.98 to 2.94 on the GPU as $N_G$ increases from 1 to 8. 
The results on the A100 confirm these findings.

With six age groups, the maximum speedups of implicit and explicit mobility, compared to serial execution on one CPU core, are 119.3 and 61.3 on 128 CPU cores, at $N_P=4096$ and $N_P=1024$ with $h=1/64$, and 493 and 331 on the GPU, both at $N_P=16384$ with $h=1/2$.
On the A100, the maximum speedups are 257.9 for implicit and 204.4 for explicit mobility.

\subsection{Parallel scalability on the CPU}
\label{sec:parallel-scalability}

We study CPU strong and weak scaling for the daily commuting scenario from~\cref{sec:runtime-comparison} with $N_G=6$ age groups, 32 simulated days, and $h=1/2$ or $h=1/64$ days. The OpenMP implementations use $n\in\{1,2,4,8,16,32,64,128\}$ threads, each of which is pinned consecutively to the physical cores. For implicit mobility, all threads access shared arrays that are spread evenly across the used cores. For explicit mobility, each thread initializes and accesses the traveler data of its patches.
Timings include all integration, synchronization, and mobility operations, including explicit departure and return. Memory allocation, initial setup, and output are excluded. All configurations are run ten times and medians are reported. 

\subsubsection{Strong scaling}
\label{sec:strong-scaling}

\begin{figure}[!t]
    \centering
    \includegraphics[width=\textwidth]{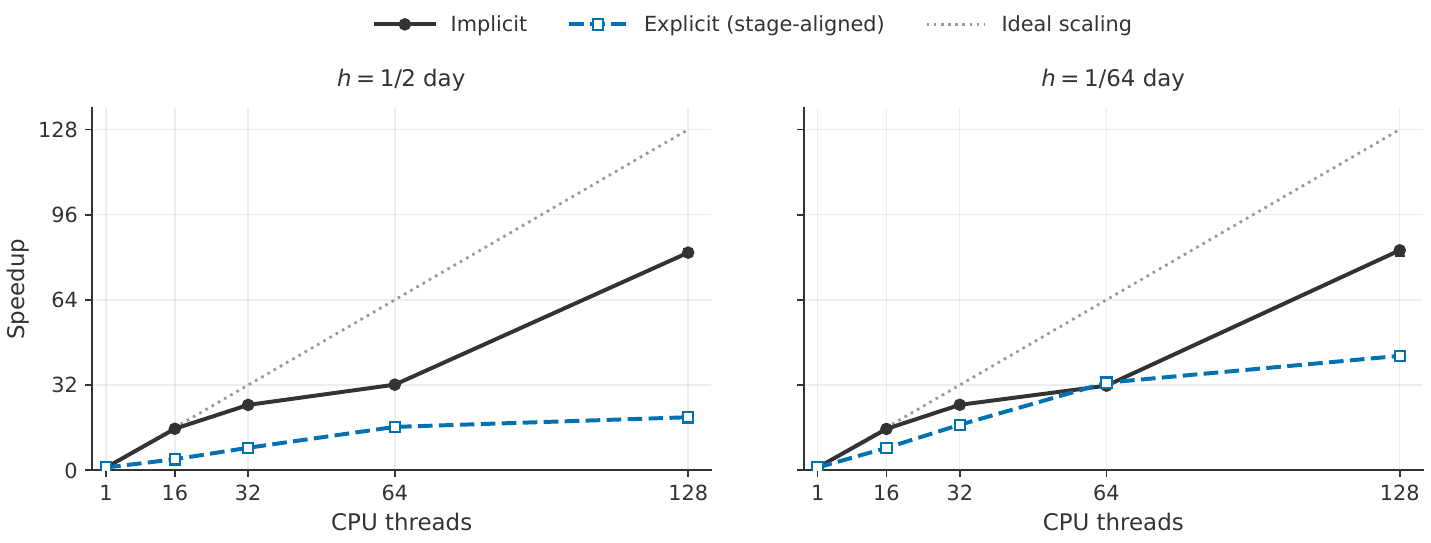}
    \caption{\textbf{CPU strong scaling with $\mathbf{N_P=8192}$ and $\mathbf{N_G=6}$.} Speedups are based on median runtimes from ten repetitions, relative to each model's one-thread OpenMP execution.}
    \label{fig:strong-scaling}
\end{figure}

For strong scaling, the problem size remains fixed at $N_P=8192$. For each model and step size, let $T(n)$ denote the median runtime of its OpenMP implementation with $n$ threads. We report speedup and parallel efficiency. 
As shown in~\cref{fig:strong-scaling}, we obtain nearly linear scaling for implicit mobility from one to eight threads. The corresponding speedup is 7.9 at both step sizes. Explicit mobility reaches a speedup of 3.5 and 5.7 at eight threads for $h=1/2$ and $h=1/64$, respectively. For 128 threads, implicit model speedups are 82.0 and 81.9, leading to efficiencies of $64.1~\%$ and $64.0~\%$. Explicit mobility reaches 20.7 and 44.0, or efficiencies of $16.1~\%$ and $34.4~\%$. With 64 threads on one CPU, the efficiencies are $50.0~\%$ and $49.6~\%$ for implicit and $25.6~\%$ and $51.3~\%$ for explicit mobility. Implicit mobility with its matrix products thus scales better overall, while explicit mobility with $h=1/64$ reaches a similar efficiency up to 64 threads.

However, better efficiency does not necessarily imply a shorter runtime. At 128 threads and $h=1/64$, explicit mobility takes 12.14~s versus 16.75~s for implicit mobility and is faster by a factor of 1.38.
The scaling curves also depend on the hardware configuration as increasing the thread count from 64 to 128 activates the second CPU, providing more bandwidth and improving implicit runtime by factors of 2.56 and 2.58, compared with 1.26 and 1.34 for explicit mobility, see the last segments in~\cref{fig:strong-scaling}.

\subsubsection{Weak scaling}
\label{sec:weak-scaling}
For weak scaling,  we let $N_P$ grow approximately with $512\sqrt{n}$ as, for dense networks, both schemes have leading quadratic work in $N_P$. We therefore consider $N_P\in\{512,724,1024,1448,2048,2896,4096,5792\}$.

\begin{figure}[!t]
    \centering
    \includegraphics[width=\textwidth]{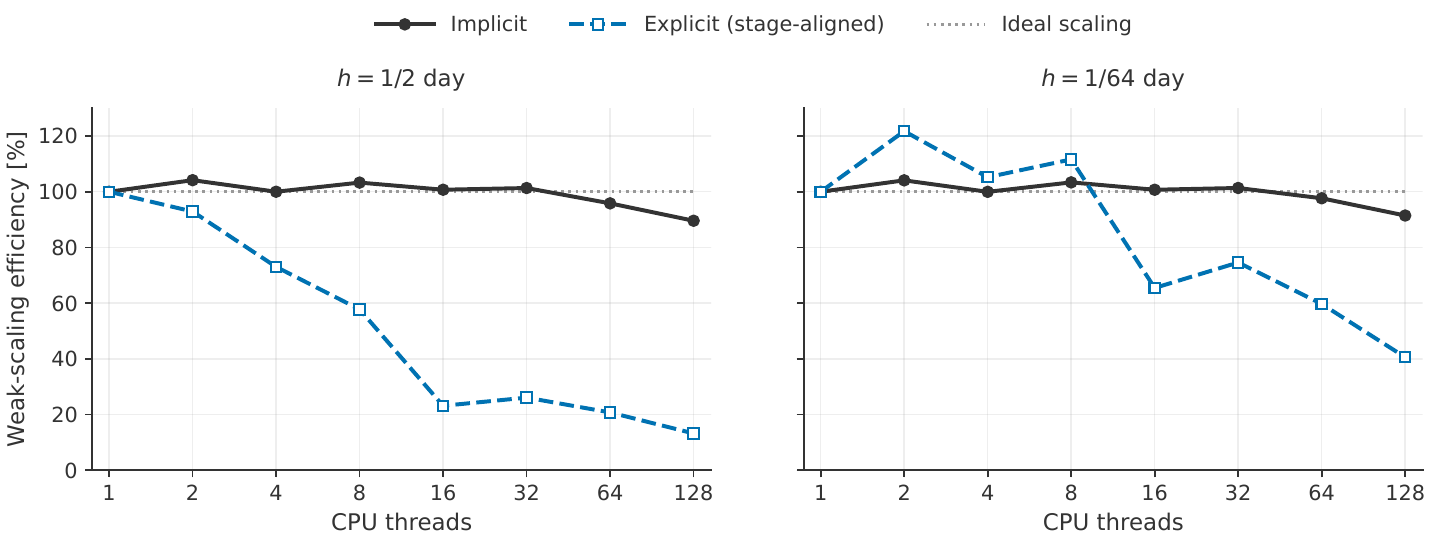}
    \caption{\textbf{CPU weak scaling with $\mathbf{N_G=6}$ and $\mathbf{N_P\approx512\sqrt{n}}$ for $\mathbf{n}$ threads}. Efficiencies are based on median runtimes from ten repetitions, relative to each model's one-thread OpenMP execution at $N_P=512$.}
    \label{fig:weak-scaling}
\end{figure}

Figure~\ref{fig:weak-scaling} shows efficiencies close to $100~\%$ up to 32 threads for implicit mobility, ending with $89.6~\%$ and $91.5~\%$ at 128 threads for $h=1/2$ and $h=1/64$, respectively. For $h=1/64$, explicit mobility keeps efficiencies around or slightly above $100~\%$ up to eight threads and reaches $59.7~\%$ at 64 threads. For $h=1/2$, the efficiency decreases to $57.7~\%$ at eight and $13.2~\%$ at 128 threads. 

\FloatBarrier

\subsubsection{Subroutine-resolved timings for parallelized explicit mobility}
\label{sec:explicit-phase-timings}

In order to quantify the cost of the different subroutines (or phases) of Algorithm~\ref{alg:stage-aligned-openmp}, we measure their runtimes summed over the entire simulation of 32 days, i.e., the home phase with the integration of the patch aggregates while all residents are at home, departure, the mobility phase with the integration of aggregates and travelers, and return. We consider strong scaling with 16 and 128 threads at $N_P=8192$ and weak scaling with 16, 32, 64, and 128 threads at $N_P\in\{2048, 2896, 4096, 5792\}$. Departure copies the resident states and then computes the traveler states and the updated aggregates. Return reconstructs the residents at home by~\cref{eq:explicit-home-reconstruction}, adds the returning travelers, and writes the new resident states back.

\begin{figure}[!t]
    \centering
    \includegraphics[width=\textwidth]{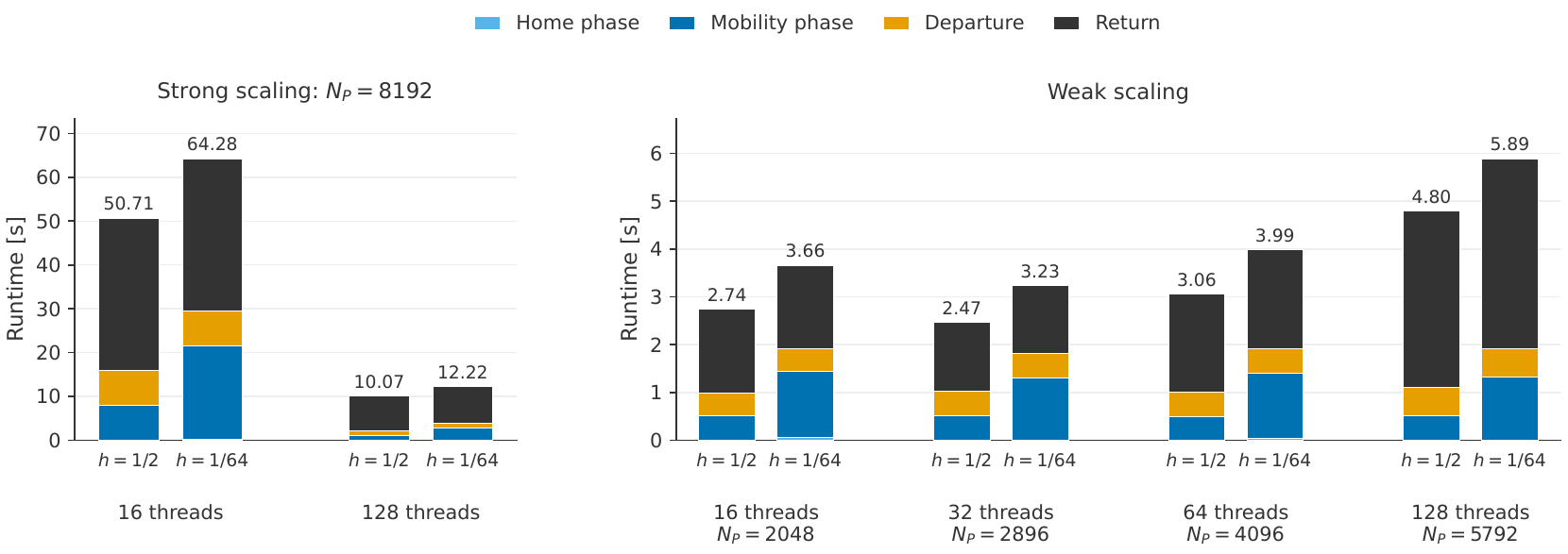}
    \caption{\textbf{CPU phase timings of the explicit 32-day simulation with $\mathbf{N_G=6}$}. Stacks show mean phase times from ten repetitions for strong scaling (left) and weak scaling (right). Step sizes $h$ are in days.}
    \label{fig:explicit-phases}
\end{figure}

\cref{fig:explicit-phases} shows mean phase times for strong (left) and weak (right) scaling. At $N_P=8192$, increasing the thread count from 16 to 128 accelerates the mobility phase by factors of 8.00 and 7.61 for $h=1/2$ and $h=1/64$, respectively. Return improves by only 4.37 and 4.21. At 128 threads, return takes 7.95 and 8.26~s, accounting for approximately $79~\%$ and $68~\%$ of total runtime. Including departure raises the mobility-event shares to $90~\%$ and $77~\%$. Thus, the modest increase in total explicit runtime with finer time steps does not mean that additional traveler updates are free of cost as their integration time rises from 1.00 to 2.80~s. Yet, mobility event costs still dominate.

For weak scaling from 16 to 128 threads, the mobility phase remains nearly constant. Return accounts for approximately $94~\%$ and $100~\%$ of the total runtime increase for $h=1/2$ and $h=1/64$, respectively. The measurements with 32 and 64 threads also identify return as the main reason why runtime decreases at 32 threads and increases again at 64 threads. For example, at $h=1/64$, return takes 1.75, 1.42, and 2.08~s at 16, 32, and 64 threads.

As described in~\cref{sec:parallel-stage-aligned}, the traveler values are stored contiguously per patch in which the travelers are located. Departure and integration therefore access consecutive memory, whereas return reads, for each home patch, one value from the traveler array of every other connected patch. Only one value per loaded cache line is needed for the current home patch, and most blocks belong to other threads, often in other NUMA domains. How costly these accesses are depends on the cache and memory placement, which changes with $N_P$ and the thread count and may explain this runtime pattern.
\FloatBarrier

\subsection{Roofline model for hardware limitations and potential}
\label{sec:performance-bottlenecks}

Eventually, we consider a roofline model~\cite{williams2009roofline} to more closely evaluate hardware limitations and potentials of our parallel implementation. Considering arithmetic throughput (operations per second) with respect to arithmetic intensity (operations per byte transferred to or from main memory) we determine whether our implementation is memory- or compute-bound and which potential is still untapped. For both CPU and GPU, we measure the second simulation day with $N_P\in\{1024,8192\}$, $N_G=6$, and $h\in\{1/2,1/64\}$ days and repeat each scenario three times.

\subsubsection{CPU}
\label{sec:cpu-roofline}

We use 128 physical cores of a node with two AMD EPYC 7702 processors. We count FP64 operations with Linux perf and measure memory traffic with LIKWID~5.2.2~\cite{treibig_likwid_2010}. As limits, we use the measured maximum floating-point performance of 4.07~TFLOP/s and memory bandwidth of 337~GB/s on 128 cores.

\begin{figure}[!t]
    \centering
    \includegraphics[width=\textwidth]{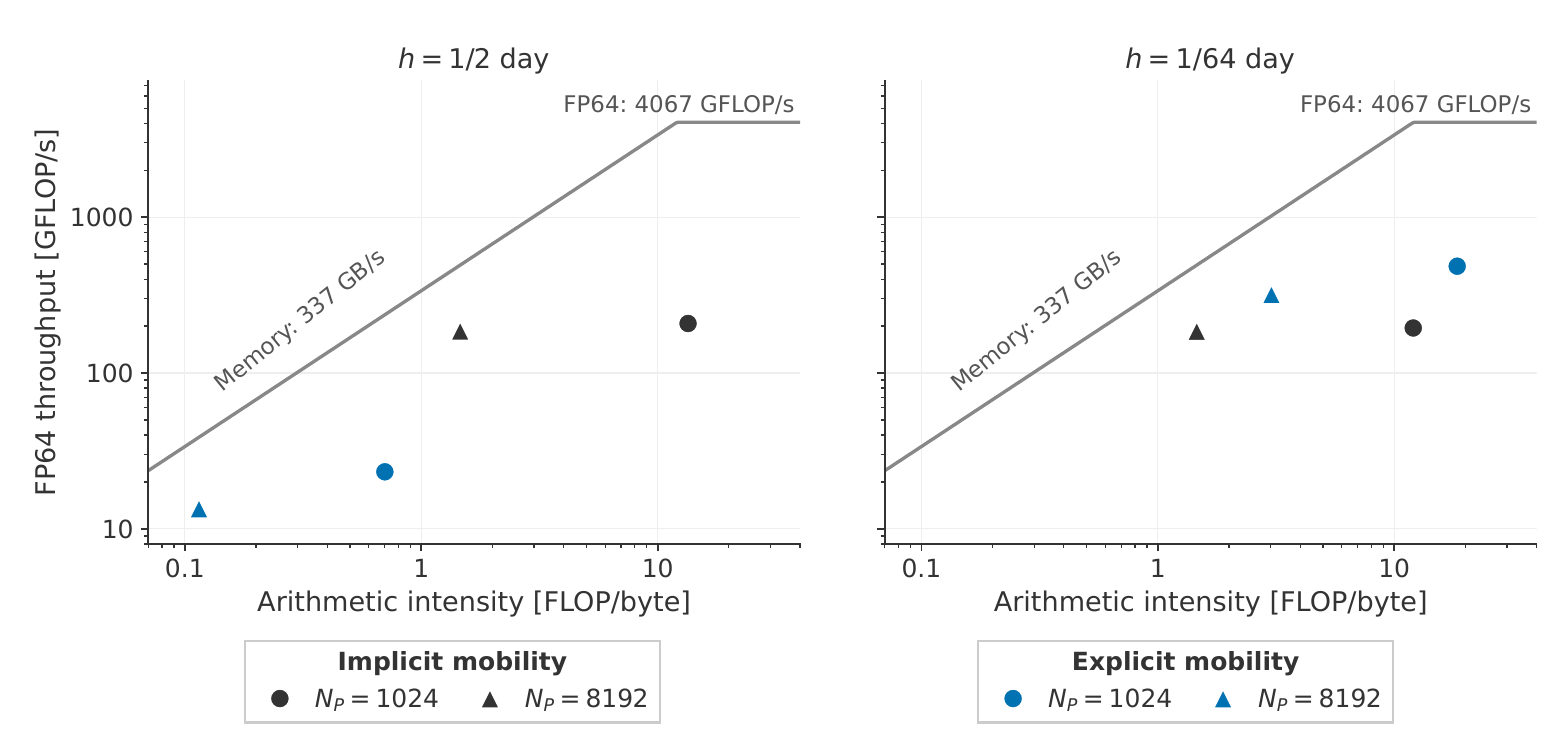}
    \caption{\textbf{CPU rooflines on 128 cores of two EPYC 7702 processors with $\mathbf{N_P\in\{1024,8192\}}$ and $\mathbf{N_G=6}$.} Markers show median results, limiting lines show measured hardware limits.}
    \label{fig:cpu-roofline}
\end{figure}

For explicit mobility,~\cref{fig:cpu-roofline} includes all four phases of~\cref{sec:explicit-phase-timings} and, for implicit mobility, all RK4 steps of the day. We calculate arithmetic intensity and throughput from the total operations, transferred bytes, and measured time, including thread synchronization and the data copies for departure and return.

At $N_P=8192$, implicit mobility reaches about 1.46~FLOP/byte and 187~GFLOP/s at both step sizes. Using $h=1/64$ increases operations, memory traffic, and time approximately 32-fold. Explicit mobility transfers about 309~GB at both step sizes, since each traveler is loaded and stored only once per mobility phase, independent of the number of steps. From $h=1/2$ to $h=1/64$, its intensity rises from 0.115 to 3.02~FLOP/byte and throughput from 13.5 to 320~GFLOP/s, while time increases only from 2.62 to 2.92~s. This result supports the benefit of reusing traveler data between mobility events.

Overall, we reach good performance with $31-38~\%$ of the reference memory bandwidth but still leaving room for improvement. The combined measurements cannot locate the bottleneck but separate phase timings in~\cref{sec:explicit-phase-timings} identified return as the main cost in the CPU scaling experiments.

\subsubsection{GPU}
\label{sec:gpu-roofline}

We measure both schemes on the NVIDIA H100 SXM 80GB using Nsight Compute~2025.1.1~\cite{nvidia_nsight_compute}, with kernels launched via CUDA Graphs as in the runtime measurements. For each kernel, we add up operations, bytes read from and written to GPU main memory, and execution times over the measured day, without the gaps between kernel launches. We then calculate throughput and arithmetic intensity and report their medians across the three runs.

The H100's Tensor Cores are specialized hardware units that accelerate matrix multiplication and accumulation.
We count all double-precision (FP64) operations, i.e., additions, multiplications, fused multiply-adds, and Tensor Core operations. Implicit matrix products use FP64 Tensor Cores, while explicit kernels use standard FP64. For H100 SXM, the theoretical maximum rates are 67 and 34~TFLOP/s, respectively, with 3.35~TB/s memory bandwidth~\cite{nvidia_h100_specifications}. Tensor Cores process fixed-size matrix blocks, so their counts also include calculations on extra entries added to fill these blocks.

\begin{figure}[!t]
    \centering
    \includegraphics[width=\textwidth]{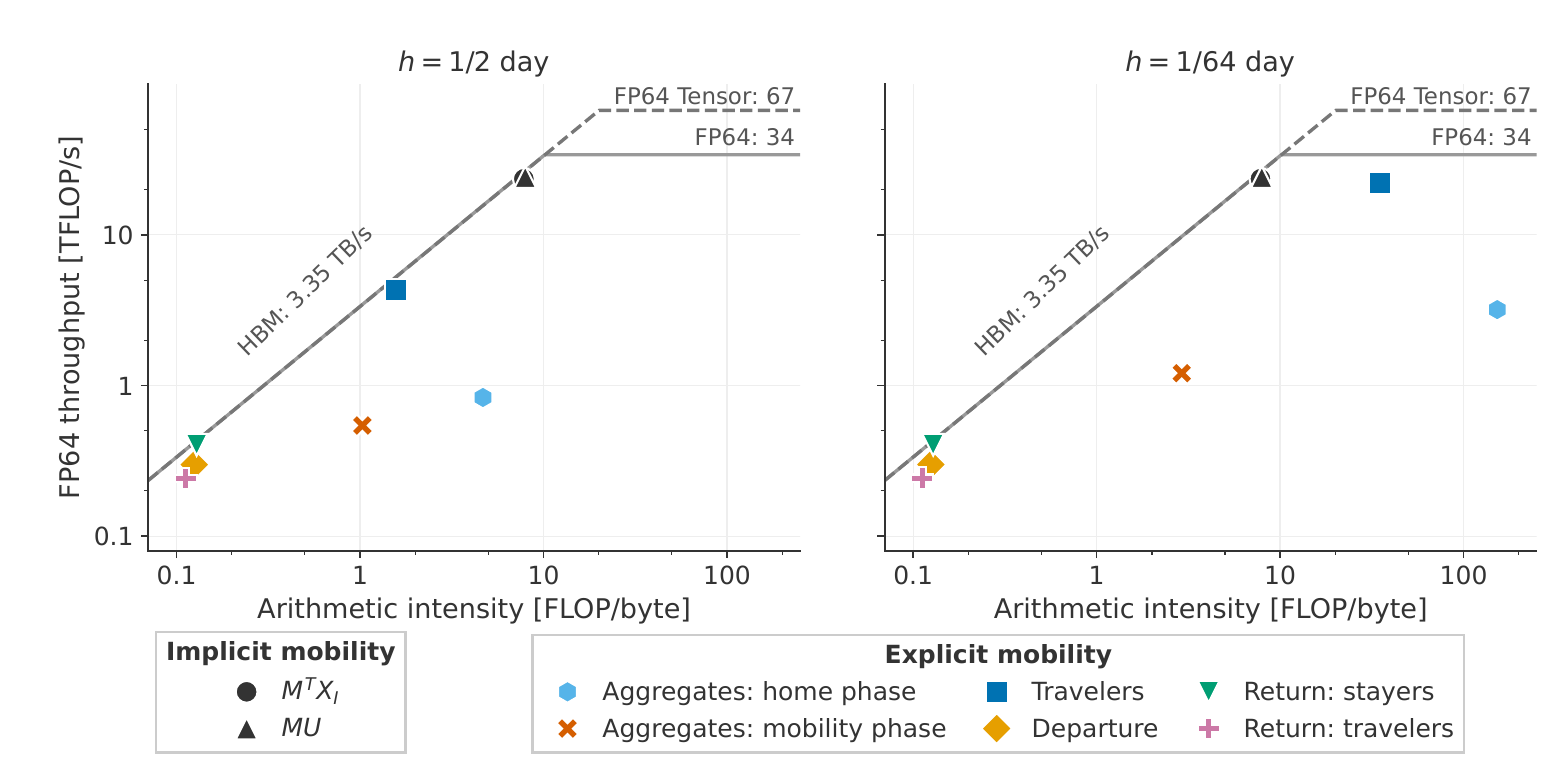}
    \caption{\textbf{GPU rooflines on NVIDIA H100 SXM with $\mathbf{N_P=8192}$ and $\mathbf{N_G=6}$.} Markers show median results, limiting lines show measured hardware limits. HBM stands for GPU's main, high-bandwidth memory. All compute kernels for the explicit model are shown. The displayed kernels for the implicit model account for more than $96~\%$ of kernel time.}
    \label{fig:gpu-roofline}
\end{figure}

\cref{fig:gpu-roofline} shows the matrix products of the implicit mobility approach and all kernels of the explicit approach. For explicit mobility, the kernels correspond to the steps of Algorithm~\ref{alg:stage-aligned-cuda}: \emph{Aggregates: home phase} and \emph{Aggregates: mobility phase} denote the aggregate kernel in these two phases, followed by \emph{Departure}, \emph{Travelers}, \emph{Return: stayers}, and \emph{Return: travelers}. The two remaining implicit kernels, which compute $U^{(s)}$ and the RK4 update, are not shown as they take only $3~\%$ of the total kernel time.

At $N_P=8192$, the matrix products in the implicit approach take approximately $97~\%$ of kernel time. Being memory-bound, they still reach about $90$-$91~\%$ of the theoretical memory-bandwidth limit. Reducing $h$ from $1/2$ to $1/64$ barely changes their positions in the plot, but increases both memory traffic and kernel time by a factor of about 32. Unsurprisingly, the established numerical linear algebra routines yield very good performance.

The explicit traveler kernel transfers about 25.8~GB at both step sizes. With $h=1/64$, it reuses the same traveler data for 32 steps between events. Its arithmetic intensity rises from 1.56 to 34.90~FLOP/byte, and its throughput reaches 22.06~TFLOP/s, or $65~\%$ of the standard FP64 maximum. Its time still increases from 9.39 to 40.84~ms. Departure and return perform only $0.11-0.13$~FLOP/byte. Together, they take $70~\%$ of explicit kernel time at $h=1/2$ and $35~\%$ at $h=1/64$. These findings support the runtime explanation in~\cref{sec:runtime-comparison}. At $N_P=1024$, $M$ fits in the H100's L2 cache, so much less data is read from main memory. The figure therefore focuses on $N_P=8192$.

\section{Conclusion}
\label{sec:conclusion}

Spatially resolved models are required to adequately represent spatio-temporal patterns and heterogeneity such as observed in large-scale epidemics or pandemics. To project the spatial spread of the disease dynamics, the inclusion of mobility connections is mandatory. In this paper, we provide a novel comparison of two major types of ODE-based metapopulation models using different ways for representing mobility. We compare an implicit mobility representation with explicit Lagrangian mobility using a recently introduced stage-aligned solution scheme~\cite{ZUNKER2027_stage_aligned}. For both approaches, we provide an extensive complexity analysis and evaluate the different contributions over increasing numbers of patches. While both approaches scale asymptotically with the numbers of edges in the mobility graph, they differ substantially in their memory-access patterns and data-reuse behavior. In addition, pre-asymptotic scaling shows substantial differences. The complexity analysis provides a first expectation on the runtime of the metapopulation models to estimate their potential to be used with increasingly higher resolution in epidemic emergencies.

In order to maximize the potential and to address the need for swift simulations that can inform decision makers, we provide optimized parallel implementations of both approaches on the CPU, using node-level parallelization through OpenMP, and on the GPU using CUDA. While the implicit mobility setting uses mobility effects in a simplified form and stores fewer population values, this approach does not automatically translate into shorter runtimes. The explicit Lagrangian mobility with discrete mobility events lets patches evolve independently between mobility events and enables a high level of parallelization. Yet, we see that with smaller mobility cycle lengths, the explicit approach converges, in a simplified setting, numerically to the implicit approach.

We found that both the age stratification and the numerical step size affect the relative performance of the two schemes and can drastically alter the performance. In particular, the numerical step size substantially changes the arithmetic intensity in the explicit mobility approach, and thus affects the performance. With 16{,}384 patches and six age groups, implicit mobility is 6.8 times faster on 128 CPU cores and 12.4 times faster on an H100 GPU when one numerical step is taken per half-day. Reducing step size to 32 steps per half-day reverses the ranking. The explicit stage-aligned formulation then runs 4.1 times faster on the CPU and 1.33 times faster on the GPU. An A100 GPU shows the same ordering, with factors of 10.4 for $h=1/2$ and 1.09 for $h=1/64$. Batching several steps allows the explicit scheme to reuse traveler data while the number of departure and return events stays fixed. The implicit formulation instead repeats its coupled matrix operations at every RK stage.

As expected, the classic numerical linear algebra based matrix operations in the implicit mobility approach yield very good weak scaling with efficiencies of about $90~\%$ and good strong scaling with efficiencies of $64~\%$ from 1 to 128 cores on a CPU. In addition, its performance on the GPU is almost on the roofline with a fully connected mobility network. The explicit mobility setting with the stage-aligned solution scheme scales less well on the CPU but still achieves a strong scaling efficiency of $51.3~\%$ when going from one to 64 CPU cores. Considering roofline models on the GPU, we see highly relevant shifts when decreasing the step size of the numerical integrator. 

Considering maximum speedup, our novel GPU implementations outperformed the corresponding serial and pre-optimized CPU implementation by factors of 493 for implicit mobility and 331 for explicit mobility. On 128 CPU cores, the matching factors are 119 and 61. Reducing selected runs from tens of minutes to just seconds highlights the importance of parallel realizations for epidemic emergencies and practical use cases.
For a synthesized pandemic application with 12{,}865 patches across the globe, we can compute approximately 10{,}000 time steps of four billion state variables in only 24~s on an H100 GPU.


\section*{Data availability}
\noindent The code to reproduce all results is available at \url{https://github.com/SciCompMod/memilio-simulations}. The population data of the worldwide application are publicly available from GPWv4~\cite{Network-CIESIN-Columbia_University_2018}.


\section*{Acknowledgements}
\noindent This work was supported by the Initiative and Networking Fund of the Helmholtz Association (grant agreement number KA1-Co-08, Project LOKI-Pandemics). It was furthermore supported by the German Federal Ministry of Education and Research and the German Federal Ministry of Research, Technology and Space under grant agreement 031L0325A (Project TwinChain) and the Deutsche Forschungsgemeinschaft (DFG, German Research Foundation) (grant agreement 528702961).

\bibliographystyle{elsarticle-num}
\bibliography{literature.bib}

@misc{godbolt,
    author = {Matt Godbolt},
    title = {Compiler explorer},
    note={\url{https://godbolt.org/}}
    }

@article{bicker_hybrid_2024,
title = {Hybrid metapopulation agent-based epidemiological models for efficient insight on the individual scale: A contribution to green computing},
journal = {Infectious Disease Modelling},
volume = {10},
number = {2},
pages = {571-590},
year = {2025},
issn = {2468-0427},
doi = {10.1016/j.idm.2024.12.015},
author = {Julia Bicker and René Schmieding and Michael Meyer-Hermann and Martin J. Kühn}
}

@article{bicker2026memilio,
	title={{MEmilio} -- A high performance Modular EpideMIcs simuLatIOn software for multi-scale and comparative simulations of infectious disease dynamics}, 
	author={Julia Bicker and Carlotta Gerstein and David Kerkmann and Sascha Korf and René Schmieding and Anna Wendler and Henrik Zunker and Daniel Abele and Maximilian Betz and Khoa Nguyen and Lena Plötzke and Kilian Volmer and Agatha Schmidt and Nils Waßmuth and Patrick Lenz and Daniel Richter and Hannah Tritzschak and Ralf Hannemann-Tamas and Julian Litz and Paul Johannssen and Marielena Borges and Annika Jungklaus and Manuel Heger and Annalena Lange and Elisabeth Kluth and Kathrin Rack and Vincent Wieland and Jonas Arruda and Sebastian Binder and Margrit Klitz and Martin Siggel and Manuel Dahmen and Achim Basermann and Michael Meyer-Hermann and Jan Hasenauer and Martin J. Kühn},
    year={2026},
	journal={Scientific Reports},
    volume = {16},    
    number={1},
    pages={26950},    
	doi={10.1038/s41598-026-66481-6}
}

@inproceedings{treibig_likwid_2010,
	address = {San Diego, CA, USA},
	title = {{LIKWID}: {A} {Lightweight} {Performance}-{Oriented} {Tool} {Suite} for x86 {Multicore} {Environments}},
	isbn = {978-1-4244-7918-4},
	shorttitle = {{LIKWID}},
	doi = {10.1109/ICPPW.2010.38},
	language = {en},
	urldate = {2023-02-06},
	booktitle = {2010 39th {International} {Conference} on {Parallel} {Processing} {Workshops}},
	publisher = {IEEE},
	author = {Treibig, Jan and Hager, Georg and Wellein, Gerhard},
	month = sep,
	year = {2010},
	pages = {207--216},
}

@inproceedings{pereira_energy_2017,
	address = {Vancouver BC Canada},
	title = {Energy efficiency across programming languages: how do energy, time, and memory relate?},
	isbn = {978-1-4503-5525-4},
	shorttitle = {Energy efficiency across programming languages},
	doi = {10.1145/3136014.3136031},
	language = {en},
	urldate = {2023-02-06},
	booktitle = {Proceedings of the 10th {ACM} {SIGPLAN} {International} {Conference} on {Software} {Language} {Engineering}},
	publisher = {ACM},
	author = {Pereira, Rui and Couto, Marco and Ribeiro, Francisco and Rua, Rui and Cunha, Jácome and Fernandes, João Paulo and Saraiva, João},
	month = oct,
	year = {2017},
	pages = {256--267},
}

@article{liu_modelling_2022,
	title = {Modelling effectiveness of {COVID}-19 pandemic control policies using an {Area}-based {SEIR} model with consideration of infection during interzonal travel},
	volume = {161},
	issn = {09658564},
	doi = {10.1016/j.tra.2022.05.003},
	language = {en},
	urldate = {2023-06-26},
	journal = {Transportation Research Part A: Policy and Practice},
	author = {Liu, Jielun and Ong, Ghim Ping and Pang, Vincent Junxiong},
	month = jul,
	year = {2022},
	pages = {25--47},
}

@article{ArinoVdD2003a,
  title = {A multi-city epidemic model},
  volume = {10},
  ISSN = {1547-724X},
  DOI = {10.1080/08898480306720},
  number = {3},
  journal = {Mathematical Population Studies},
  publisher = {Informa UK Limited},
  author = {Arino,  Julien and van den Driessche,  P.},
  year = {2003},
  month = jan,
  pages = {175–193}
}

@article{
Citron2021,
author = {Daniel T. Citron  and Carlos A. Guerra  and Andrew J. Dolgert  and Sean L. Wu  and John M. Henry  and Héctor M. {Sánchez C.}  and David L. Smith },
title = {Comparing metapopulation dynamics of infectious diseases under different models of human movement},
journal = {Proceedings of the National Academy of Sciences},
volume = {118},
number = {18},
pages = {e2007488118},
year = {2021},
doi = {10.1073/pnas.2007488118}}

@article{wesolowski2012quantifying,
  title={Quantifying the impact of human mobility on malaria},
  author={Wesolowski, Amy and Eagle, Nathan and Tatem, Andrew J and Smith, David L and Noor, Abdisalan M and Snow, Robert W and Buckee, Caroline O},
  journal={Science},
  volume={338},
  number={6104},
  pages={267--270},
  year={2012},
  publisher={American Association for the Advancement of Science},
  doi={10.1126/science.1223467}
}

@article{balcan2009multiscale,
  title={Multiscale mobility networks and the spatial spreading of infectious diseases},
  author={Balcan, Duygu and Colizza, Vittoria and Gon{\c{c}}alves, Bruno and Hu, Hao and Ramasco, Jos{\'e} J and Vespignani, Alessandro},
  journal={Proceedings of the National Academy of Sciences},
  volume={106},
  number={51},
  pages={21484--21489},
  year={2009},
  publisher={National Acad Sciences},
  doi={10.1073/pnas.0906910106}
}

@article{bajardi2011human,
  title={Human mobility networks, travel restrictions, and the global spread of 2009 {H1N1} pandemic},
  author={Bajardi, Paolo and Poletto, Chiara and Balcan, Duygu and Hu, Hao and Gon{\c{c}}alves, Bruno and Ramasco, Jos{\'e} J and Paolotti, Daniela and Perra, Nicola and Tizzoni, Michele and Van den Broeck, Wouter and others},
  journal={PLoS One},
  volume={6},
  number={1},
  pages={e16591},
  year={2011},
  publisher={Public Library of Science},
  doi={10.1371/journal.pone.0016591}
}

@article{kraemer2020mobility,
  author  = {Kraemer, Moritz U. G. and Yang, Chia-Hua and Gutierrez, Bernardo and et al.},
  title   = {The effect of human mobility and control measures on the {COVID}-19 epidemic in China},
  journal = {Science},
  year    = {2020},
  volume  = {368},
  number  = {6490},
  pages   = {493--497},
  doi     = {10.1126/science.abb4218}
}

@article{COSNER2009550,
title = {The effects of human movement on the persistence of vector-borne diseases},
journal = {Journal of Theoretical Biology},
volume = {258},
number = {4},
pages = {550-560},
year = {2009},
issn = {0022-5193},
doi = {10.1016/j.jtbi.2009.02.016},
author = {C. Cosner and J.C. Beier and R.S. Cantrell and D. Impoinvil and L. Kapitanski and M.D. Potts and A. Troyo and S. Ruan}
}

@article{
doi:10.1073/pnas.0510525103,
author = {Vittoria Colizza  and Alain Barrat  and Marc Barthélemy  and Alessandro Vespignani },
title = {The role of the airline transportation network in the prediction and predictability of global epidemics},
journal = {Proceedings of the National Academy of Sciences},
volume = {103},
number = {7},
pages = {2015-2020},
year = {2006},
doi = {10.1073/pnas.0510525103}}

@article{PhysRevX.1.011001,
  title = {Natural Human Mobility Patterns and Spatial Spread of Infectious Diseases},
  author = {Belik, Vitaly and Geisel, Theo and Brockmann, Dirk},
  journal = {Phys. Rev. X},
  volume = {1},
  issue = {1},
  pages = {011001},
  numpages = {5},
  year = {2011},
  month = {Aug},
  publisher = {American Physical Society},
  doi = {10.1103/PhysRevX.1.011001},
}

@article{Arino2015,
  title = {Epidemiological implications of mobility between a large urban centre and smaller satellite cities},
  volume = {71},
  ISSN = {1432-1416},
  DOI = {10.1007/s00285-014-0854-z},
  number = {5},
  journal = {Journal of Mathematical Biology},
  publisher = {Springer Science and Business Media LLC},
  author = {Arino,  Julien and Portet,  Stéphanie},
  year = {2015},
  month = jan,
  pages = {1243–1265}
}

@article{Rapaport2022,
  title = {The Role of Permanently Resident Populations in the Two-Patches {SIR} Model with Commuters},
  volume = {85},
  ISSN = {1522-9602},
  DOI = {10.1007/s11538-022-01111-6},
  number = {1},
  journal = {Bulletin of Mathematical Biology},
  publisher = {Springer Science and Business Media LLC},
  author = {Rapaport,  Alain and Mimouni,  Ismail},
  year = {2022},
  month = dec 
}

@inproceedings{wang2018,
author = {Wang, Jingyuan and Wang, Xiaojian and Wu, Junjie},
title = {Inferring Metapopulation Propagation Network for Intra-city Epidemic Control and Prevention},
year = {2018},
isbn = {9781450355520},
publisher = {Association for Computing Machinery},
address = {New York, NY, USA},
doi = {10.1145/3219819.3219865},
booktitle = {Proceedings of the 24th ACM SIGKDD International Conference on Knowledge Discovery \& Data Mining},
pages = {830–838},
numpages = {9},
location = {London, United Kingdom},
series = {KDD '18}
}

@article{JSSv091i12,
 title={{SimInf: An R} Package for Data-Driven Stochastic Disease Spread Simulations},
 volume={91},
 doi={10.18637/jss.v091.i12},
 number={12},
 journal={Journal of Statistical Software},
 author={Widgren, Stefan and Bauer, Pavol and Eriksson, Robin and Engblom, Stefan},
 year={2019},
 pages={1–42}
}

@article{williams2009roofline,
author = {Williams, Samuel and Waterman, Andrew and Patterson, David},
title = {Roofline: an insightful visual performance model for multicore architectures},
year = {2009},
issue_date = {April 2009},
publisher = {Association for Computing Machinery},
address = {New York, NY, USA},
volume = {52},
number = {4},
issn = {0001-0782},
doi = {10.1145/1498765.1498785},
journal = {Commun. ACM},
month = apr,
pages = {65–76},
numpages = {12}
}

@article{ZUNKER2027_stage_aligned,
title = {Efficient numerical computation of traveler states in explicit mobility-based metapopulation models: Mathematical theory and application to epidemics},
journal = {Mathematics and Computers in Simulation},
volume = {251},
pages = {798-812},
year = {2027},
issn = {0378-4754},
doi = {10.1016/j.matcom.2026.08.029},
author = {Henrik Zunker and René Schmieding and Jan Hasenauer and Martin J. Kühn},
}

@misc{nvidia_h100_specifications,
  author       = {{NVIDIA}},
  title        = {{NVIDIA H100 Tensor Core GPU}: Specifications},
  year         = {n.d.},
  howpublished = {\url{https://www.nvidia.com/en-us/data-center/h100/}},
  note         = {Accessed 25 September 2026}
}

@article{contreras2026mechanics,
  title = {Mechanics of pandemics},
  volume = {98},
  ISSN = {2589-5370},
  DOI = {10.1016/j.eclinm.2026.104101},
  journal = {eClinicalMedicine},
  publisher = {Elsevier BV},
  author = {Contreras,  Seba and D\"{o}nges,  Philipp and M\"{u}ller,  Laura and Mallick,  Piklu and Paltra,  Sydney and Hvid,  Ulrik and Kettlitz,  Robyn J.N. and Reitenbach,  Andreas and Lind,  Rodrigo Amaral and Aguiar,  Maíra and Bechtle,  Philip and Valdez,  André Calero and Gronemeyer,  Ronja and Harries,  Manuela and Jaeger,  Veronika K. and Karch,  André and Klett-Tammen,  Carolina J. and Klimek,  Peter and Kretzschmar,  Mirjam E. and Nagel,  Kai and Nielsen,  Bjarke Frost and Prainsack,  Barbara and Radhuber,  Isabella M. and Simonsen,  Lone and Sneppen,  Kim and Suer,  Janik and Priesemann,  Viola},
  year = {2026},
  month = Aug,
  pages = {104101}
}

@article{Kovac2018,
  title = {Heterogeneous computing for epidemiological model fitting and simulation},
  volume = {19},
  ISSN = {1471-2105},
  DOI = {10.1186/s12859-018-2108-3},
  number = {1},
  journal = {BMC Bioinformatics},
  publisher = {Springer Science and Business Media LLC},
  author = {Kovac,  Thomas and Haber,  Tom and Van Reeth,  Frank and Hens,  Niel},
  year = {2018},
  month = Mar 
}

@article{kehrer_hybrid_2027,
	title = {A hybrid {ABM}-{PDE} framework for real-world infectious disease simulations},
	volume = {162},
	issn = {0307-904X},
	doi = {10.1016/j.apm.2026.117173},
	urldate = {2026-08-11},
	journal = {Applied Mathematical Modelling},
	author = {Kehrer, Kristina and Conrad, Tim O. F.},
	month = feb,
	year = {2027},
	pages = {117173},
}

@article{perumalla2021mesoscopic,
  title = {Mesoscopic Modeling and Rapid Simulation of Incremental Changes in Epidemic Scenarios on {GPUs}: Fast What–If Analyses of Localized and Dynamic Effects},
  volume = {101},
  ISSN = {0019-4964},
  DOI = {10.1007/s41745-021-00253-1},
  number = {3},
  journal = {Journal of the Indian Institute of Science},
  publisher = {Springer Science and Business Media LLC},
  author = {Perumalla,  Kalyan S. and Alam,  Maksudul},
  year = {2021},
  month = July,
  pages = {357–370}
}

@misc{irdr_peril,
    author = {{Integrated Research on Disaster Risk}},
    year = {2014},
    title = {Peril Classification and Hazard Glossary ({IRDR DATA} Publication No. 1)},
    note = {\url{https://www.irdrinternational.org/knowledge_pool/publications/13}}
}

@article{tonnelier_what_2025,
	title = {What makes an epidemic a disaster: the future of epidemics within the {EM}-{DAT} {International} {Disaster} {Database}},
	volume = {25},
	issn = {1471-2458},
	shorttitle = {What makes an epidemic a disaster},
	doi = {10.1186/s12889-024-21026-2},
	urldate = {2026-10-15},
	journal = {BMC Public Health},
	author = {Tonnelier, Margo and Delforge, Damien and Below, Regina and Munguía, Juan Armando Torres and Saegerman, Claude and Wathelet, Valentin and van Loenhout, Joris A. F. and Speybroeck, Niko},
	month = jan,
	year = {2025},
	pages = {21},
}

@article{bagaforo_scoping_2026,
	title = {A scoping review of {COVID}-19 modelling studies in {Belgium} 2020-2024: incorporation of behaviour and lessons learned},
	volume = {84},
	issn = {2049-3258},
	shorttitle = {A scoping review of {COVID}-19 modelling studies in {Belgium} 2020-2024},
	doi = {10.1186/s13690-026-01959-3},
	language = {en},
	number = {1},
	urldate = {2026-10-15},
	journal = {Archives of Public Health},
	author = {Bagaforo, Roel Jude and Dupas, Marie-Cécile and Abrams, Steven and Dellicour, Simon and Hens, Niel},
	month = may,
	year = {2026},
	pages = {160},
}

@techreport{ecdc_npis_2024,
	institution ={European Centre for Disease Prevention and Control},
	title = {Public health and social measures for health emergencies and pandemics in the {EU}/{EEA}: recommendations for strengthening preparedness planning},
	shorttitle = {Public health and social measures for health emergencies and pandemics in the {EU}/{EEA}},
	language = {en},
	urldate = {2024-09-29},
	author = {{European Centre for Disease Prevention and Control}},
	year = {2024},
	doi = {10.2900/253991}
}

@mastersthesis{gerstein_MA,
            year = {2025},
           title = {Evaluation of different mobility models in equation-based infectious disease dynamics},
          author = {Gerstein, Carlotta},
             note = {\url{https://elib.dlr.de/213667/}},
             school = {University of Cologne}
}

@article{dagum1998openmp,
  author  = {Dagum, L. and Menon, R.},
  title   = {{OpenMP}: An Industry Standard {API} for
             Shared-Memory Programming},
  journal = {IEEE Computational Science and Engineering},
  year    = {1998},
  volume  = {5},
  number  = {1},
  pages   = {46--55},
  doi     = {10.1109/99.660313}
}

@article{nickolls2008cuda,
  author  = {Nickolls, John and Buck, Ian and Garland, Michael
             and Skadron, Kevin},
  title   = {Scalable Parallel Programming with {CUDA}},
  journal = {Queue},
  year    = {2008},
  volume  = {6},
  number  = {2},
  pages   = {40--53},
  doi     = {10.1145/1365490.1365500}
}

@misc{nvidia_nsight_compute,
  author       = {{NVIDIA}},
  title        = {{Nsight Compute}},
  year         = {2025},
  howpublished = {\url{https://developer.nvidia.com/nsight-compute}},
  note         = {Version 2025.1.1}
}

@misc{Network-CIESIN-Columbia_University_2018,
  title={Gridded Population of the World, Version 4 ({GPWv4}): Population Count, Revision 11},
  DOI={10.7927/H4JW8BX5},
  publisher={Palisades, NY: NASA Socioeconomic Data and Applications Center (SEDAC)},
  author={{Center for International Earth Science Information Network - CIESIN - Columbia University}},
  year={2018},
}

@article{Barbosa_2018,
   title={Human mobility: Models and applications},
   volume={734},
   ISSN={0370-1573},
   DOI={10.1016/j.physrep.2018.01.001},
   journal={Physics Reports},
   publisher={Elsevier BV},
   author={Barbosa, Hugo and Barthelemy, Marc and Ghoshal, Gourab and James, Charlotte R. and Lenormand, Maxime and Louail, Thomas and Menezes, Ronaldo and Ramasco, José J. and Simini, Filippo and Tomasini, Marcello},
   year={2018},
   month=Mar, pages={1–74} }

\newpage
\appendix
\setcounter{table}{0}
\pagenumbering{roman}

\section{Operation counts of the complexity terms}
\label{app:operation-counts}

We consider local SEIR models ($N_C=4N_G$, $N_T=3N_G$) coupled through a fully connected network ($N_P+N_E=N_P^2$ entries in $M$). Since only the infectious compartments are included in the infection rates, the aggregations (i) and (iii) of the implicit mobility only affect $N_G$ of the $N_C$ compartments and the $N_T$ transitions, respectively. In particular, the intensity evaluation in (ii), $\mathcal{C}_D(N_C,N_T)$, grows with $N_G^2$.

\begin{table}[H]
\centering
\small
\renewcommand{\arraystretch}{1.3}
\begin{tabularx}{\linewidth}{@{}l >{\raggedright\arraybackslash}p{3.9cm} l >{\raggedright\arraybackslash}X@{}}
\toprule
Term & Operation & Prefactor & Derivation \\
\midrule
(i) & Aggregation $M^TX_I$, divided by $\widetilde N$ & $2N_G(N_P+N_E)+N_GN_P$ & 2 operations per entry of $M$ and age group to multiply by $I$ and accumulate, and 1 operation per age group to multiply by $1/\widetilde N$. \\
(ii) & Intensity evaluation & $(3N_G^2+N_G)N_P$ & 1 operation per age group for the home share $I/N$, and 3 operations per pair of age groups to add the home and mobile shares, multiply by the contact coefficient, and accumulate. \\
(iii) & Intensity aggregation $MU$ & $2N_G(N_P+N_E)$ & 2 operations per entry of $M$ and age group to multiply by $U$ and accumulate. \\
(iv) & Right-hand side & $5N_GN_P$ & 3 operations per age group for the flows $S\to E$, $E\to I$, and $I\to R$ (one multiplication each) and 2 operations for the derivatives of $E$ and $I$ (one difference each). \\
(v) & RK stage combinations & $10.5N_GN_P$ & 4 operations per compartment $S$, $E$, and $I$ in stages 1--3 to update the accumulator and the next stage state (one multiplication and one addition each), and 2 operations in stage 4 for the accumulator only, i.e., 3.5 operations on average. $R$ follows from the conserved population. \\
\bottomrule
\end{tabularx}
\caption{\textbf{Floating point operations per RK4 stage of the terms in~\cref{eq:implicit-complexity} for implicit mobility.}}
\label{tab:operation-counts-implicit}
\end{table}

\begin{table}[H]
\centering
\small
\renewcommand{\arraystretch}{1.3}
\begin{tabularx}{\linewidth}{@{}l >{\raggedright\arraybackslash}p{3.9cm} l >{\raggedright\arraybackslash}X@{}}
\toprule
Term & Operation & Prefactor & Derivation \\
\midrule
(i) & Intensity evaluation per aggregate & $3N_G^2N_P$ & 3 operations per pair of age groups to compute $\beta\cdot I\cdot(1/N)$ and accumulate. \\
(ii) & Right-hand side of the aggregates & $5N_GN_P$ & As (iv) for implicit mobility. \\
(iii) & Right-hand side of the travelers & $5N_GN_E$ & As (ii), for each traveling group. \\
(iv) & RK stage combinations & $12N_GN_E+12.25N_GN_P$ & 10.5 as (v) for implicit mobility. In addition, 6 operations per step, i.e., 1.5 per stage, for $N$ as a sum (3 additions) and $R$ from the conserved population (3 subtractions). The aggregates additionally require 1 operation per step, i.e., 0.25 per stage, for $1/N$ (one division). \\
\midrule
Events & Departure and return, per day & $16N_GN_E$ & 4 operations per compartment: at departure, computing the travelers (fraction $m_{pq}$ times residents, one multiplication) and adding them to the aggregate, and at return, subtracting them from the aggregate and adding them to the home patch. \\
\bottomrule
\end{tabularx}
\caption{\textbf{Floating point operations per RK4 stage of the terms in~\cref{eq:stage-aligned-complexity} for the mobility phase of explicit mobility, and operations of the mobility events per day.}}
\label{tab:operation-counts-explicit}
\end{table}

\section{Complexity analysis of the direct global integration}
\label{sec:direct-global-complexity}

For a better comparison with the stage-aligned scheme, we here also provide the complexity analysis for the direct global integration approach.

\subsection{Arithmetic work}

Direct global integration applies the $S$-stage RK method jointly to all $(N_P+N_E)N_C$ origin-destination variables described by~\cref{eq:explicit-model-standard}. At each stage, forming the patch aggregates (i) according to~\cref{eq:explicit-populations} costs $\mathcal{O}((N_P+N_E)N_C)$. This additional pass over all origin-destination states is required even if transition intensities are subsequently reused.
The remaining computational cost depends on whether the direct implementation reevaluates (see~\cref{eq:direct-global-complexity-recompute}) or reuses once calculated (see~\cref{eq:direct-global-complexity-reuse}) transition intensities for every origin-specific population located in the same patch.

A straightforward implementation evaluates the right-hand side in~\cref{eq:explicit-model-standard} separately for every $(p,q)\in\mathcal{E}_+$, reevaluating the transition-intensity matrix $D^{(q)}(\widetilde{x}^{(q)},t)$ for each origin-specific population located in $q$. (ii) Therefore, the intensity evaluations cost $\mathcal{O}((N_P+N_E)\mathcal{C}_D(N_C,N_T))$ per stage. (iii) Computing the right-hand side, after the intensities are known, costs $\mathcal{O}((N_P+N_E)(N_C+N_T))$ per stage, and (iv) the RK combinations cost $\mathcal{O}(S^2(N_P+N_E)N_C)$ over the complete step. Hence,
\begin{align}
    \mathcal{W}_{\mathrm{ex,direct}}^{\mathrm{recompute}}
    &=\mathcal{O}\left(
        \begin{aligned}
        \underbrace{S(N_P+N_E)N_C}_{\text{(i)}}
        +\underbrace{S(N_P+N_E)\mathcal{C}_D(N_C,N_T)}_{\text{(ii)}}
        +\underbrace{S(N_P+N_E)(N_C+N_T)}_{\text{(iii)}}
        +\underbrace{S^2(N_P+N_E)N_C}_{\text{(iv)}}
        \end{aligned}
    \right).
    \label{eq:direct-global-complexity-recompute}
\end{align}

For fixed $S$, this reduces to $\mathcal{O}((N_P+N_E)(\mathcal{C}_D(N_C,N_T)+N_C+N_T))$.

The redundant intensity evaluations can be removed by evaluating $D^{(q)}(\widetilde{x}^{(q)},t)$ once per patch and reusing it for all origin-specific populations currently located in $q$. This custom patch-wise organization reduces (ii) the number of intensity evaluations per stage from $N_P+N_E$ to $N_P$, but (i) the stage-wise aggregation, (iii) origin-specific right-hand side evaluations, and (iv) global RK combinations remain unchanged:
\begin{align}
    \mathcal{W}_{\mathrm{ex,direct}}^{\mathrm{reuse}}
    &=\mathcal{O}\left(
        \begin{aligned}
        \underbrace{S(N_P+N_E)N_C}_{\text{(i)}}
        +\underbrace{SN_P\mathcal{C}_D(N_C,N_T)}_{\text{(ii)}}
        +\underbrace{S(N_P+N_E)(N_C+N_T)}_{\text{(iii)}}
        +\underbrace{S^2(N_P+N_E)N_C}_{\text{(iv)}}
        \end{aligned}
    \right).
    \label{eq:direct-global-complexity-reuse}
\end{align}
For fixed $S$,~\cref{eq:direct-global-complexity-reuse} reduces to $\mathcal{O}(N_P\mathcal{C}_D(N_C,N_T)+(N_P+N_E)(N_C+N_T))$. This simplified bound includes the $\mathcal{O}((N_P+N_E)N_C)$ operations needed to rebuild the patch aggregates at every RK stage.

\subsection{Memory usage}

Both realizations hold (i) $(N_P+N_E)N_C$ origin-destination values, (ii) $N_PN_C$ patch-aggregate values, and (iv) $S(N_P+N_E)N_C$ RK stage-derivative values. (iii) Each equation-wise intensity evaluation requires only one $N_T$-value temporary array and (v) mobility coefficients require $\mathcal{O}(N_P+N_E)$ values:
\begin{align}
    \mathcal{M}_{\mathrm{ex,direct}}^{\mathrm{recompute}}
    &=\mathcal{O}\left(
        \begin{aligned}
        \underbrace{(N_P+N_E)N_C}_{\text{(i)}}
        +\underbrace{N_PN_C}_{\text{(ii)}}
        +\underbrace{N_T}_{\text{(iii)}}
        +\underbrace{S(N_P+N_E)N_C}_{\text{(iv)}}
        +\underbrace{N_P+N_E}_{\text{(v)}}
        \end{aligned}
    \right)
    \label{eq:direct-global-memory-recompute}
\end{align}
This simplifies to $\mathcal{M}_{\mathrm{ex,direct}}^{\mathrm{recompute}}=\mathcal{O}(S(N_P+N_E)N_C+N_T)$.

Patch-wise reuse has the same memory bound if patches are processed one at a time and the $N_T$-value temporary array is reused. If the intensities of all patches are instead stored for later access, the additional $N_PN_T$-value table gives
\begin{align}
    \mathcal{M}_{\mathrm{ex,direct}}^{\mathrm{reuse,table}}
    &=\mathcal{O}\left(S(N_P+N_E)N_C+N_PN_T\right).
    \label{eq:direct-global-memory-reuse-table}
\end{align}
A reduced-storage RK implementation, which does not require to keep all $S$ stage derivatives at the same time, can reduce the stage-derivative term, but it does not reduce the $(N_P+N_E)N_C$ global state or eliminate the stage-wise aggregation.

\newpage
\section{Model and simulation setup}
\setcounter{table}{0}
\label{app:mobility-outcome-setup}

\begin{table}[htbp]
\small
\centering
\caption{\textbf{Model and setup parameters used for the comparison in~\cref{sec:model-outcomes}}. All epidemiological parameters are constant in time and identical across patches; the mobility matrix is also constant.}
\label{tab:mobility-outcome-setup}
\setlength{\tabcolsep}{4pt}
\renewcommand{\arraystretch}{1.08}
\begin{tabularx}{\textwidth}{@{}L{1.8cm}L{4.2cm}X@{}}
\toprule
\textbf{Symbol} & \textbf{Description} & \textbf{Setting} \\
\midrule
$N_P$ & Number of patches & 4 \\
$N_G$ & Number of age groups & 1 (hence $N_C=4$ and $N_T=3$) \\
$N^{(p)}$ & Resident population & $10{,}000$ in every patch \\
$x^{(p)}(0)$ & Initial state & $(S,E,I,R)=(9900,80,20,0)$ in Patch~1 and $(10{,}000,0,0,0)$ in Patches~$2-4$.\\
\midrule
$\rho$ & Infection risk per contact & $0.05$ \\
$\phi$ & Mean daily contacts & $6$ \\
$T_E$ & Mean exposed period & $4$ days \\
$T_I$ & Mean infectious period & $6$ days \\
\midrule
$M$ & Mobility matrix & $m_{11}=m_{44}=0.9$, $m_{22}=m_{33}=0.8$, and $m_{p,p+1}=m_{p+1,p}=0.1$ for $p=1,2,3$; all other entries are zero. \\
$\theta$ & Weight of the home phase & $1/2$ \\
$\delta$ & Explicit mobility-cycle length & $\{0.25,0.5,1,2,4\}$ days; $\delta=1$ day is the primary case \\
$\tau_a$ & Mobility events & Departure at $\tau_{2a+1}=a\delta+\delta/2$ and return at $\tau_{2a+2}=(a+1)\delta$, where $a\in\mathbb{N}_0$. \\
$m_a$ & Event fractions & For $(p,q)\in\mathcal{E}$, departure uses $m_a^{(p;p\to q)}=m_{pq}$ and $m_a^{(p;q\to p)}=0$. Return uses $m_a^{(p;q\to p)}=1$ and $m_a^{(p;p\to q)}=0$. \\
\midrule
$t_{\max}$ & Simulation horizon & $160$ days \\
$h$ & Integration step size & $1/64$ day \\ 
\bottomrule
\end{tabularx}
\end{table}

\begin{table}[htbp]
\small
\centering
\caption{\textbf{Model and setup parameters used for the runtime comparison in~\cref{sec:runtime-comparison}}. Patches are indexed by $p,q=1,\ldots,N_P$ and age groups by $i,j=1,\ldots,N_G$. All epidemiological parameters are constant in time and identical across patches; the mobility matrix is also constant.}
\label{tab:runtime-setup}
\setlength{\tabcolsep}{4pt}
\renewcommand{\arraystretch}{1.08}
\begin{tabularx}{\textwidth}{@{}L{1.8cm}L{4.2cm}X@{}}
\toprule
\textbf{Symbol} & \textbf{Description} & \textbf{Setting} \\
\midrule
$N_i^{(p)}$ & Resident population & $10^4\,\bigl(0.9+0.2\,p/(N_P+1)\bigr)\bigl(0.9+0.2\,i/(N_G+1)\bigr)/N_G$ \\
$x_i^{(p)}(0)$ & Initial state & $E_i^{(p)}=N_i^{(p)}\bigl(0.007+0.003\,p/(N_P+1)\bigr)/\bigl(1+0.1(i-1)\bigr)$, $I_i^{(p)}=N_i^{(p)}\bigl(0.012-0.004\,p/(N_P+1)\bigr)\bigl(1+0.08(i-1)\bigr)$, $R_i^{(p)}=N_i^{(p)}\bigl(0.02+0.002(i-1)\bigr)$, and $S_i^{(p)}=N_i^{(p)}-E_i^{(p)}-I_i^{(p)}-R_i^{(p)}$ \\
\midrule
$\rho_i$ & Infection risk per contact & $0.035+0.002(i-1)$ \\
$\phi_{ij}$ & Mean daily contacts & $9\bigl(1+0.02(i-1)+0.01(j-1)\bigr)/\bigl(1+|i-j|\bigr)$ \\
$T_{E_i}$ & Mean exposed period & $5.2+0.1(i-1)$ days \\
$T_{I_i}$ & Mean infectious period & $6+0.1(i-1)$ days \\
\midrule
$M$ & Mobility matrix & $m_{pp}=1-\mu_p$ with commuting fraction $\mu_p=0.1+0.1\bigl((p-1)\bmod 7\bigr)/6$, and $m_{pq}=\mu_p w_{pq}/\sum_{r\neq p}w_{pr}$ for $q\neq p$ with $w_{pq}=1+0.02\bigl((17p+13q-30)\bmod 23\bigr)$ \\
$\theta$ & Weight of the home phase & $1/2$ \\
$\tau_a$ & Mobility events & Departure at $\tau_{2a+1}=a+1/2$ and return at $\tau_{2a+2}=a+1$~days, where $a\in\mathbb{N}_0$ \\
\midrule
$h$ & Integration step size & $1/2$ or $1/64$~day, i.e., one or 32 steps per half-day \\
\bottomrule
\end{tabularx}
\end{table}
\FloatBarrier

\newpage
\section{Butcher tableau of the numerical solver}
\setcounter{table}{0}
\label{app:rk-tableaus}

All simulations use the classical explicit fourth-order Runge-Kutta method with the following Butcher tableau $(c, A, b^T)$:
\begin{table}[!h]
    \centering
    \begin{minipage}{0.45\textwidth}
        \centering
        \begin{tabular}{c|cccc}
            0   & 0   & 0   & 0 & 0 \\
            $\frac{1}{2}$ & $\frac{1}{2}$ & 0 & 0 & 0 \\
            $\frac{1}{2}$ & 0 & $\frac{1}{2}$ & 0 & 0 \\
            1   & 0   & 0   & 1 & 0 \\
            \hline
                & $\frac{1}{6}$ & $\frac{1}{3}$ & $\frac{1}{3}$ & $\frac{1}{6}$
        \end{tabular}
    \end{minipage}
\end{table}
\end{document}